\documentstyle[preprint,aps,floats,psfig]{revtex}
\tighten
\begin{document}
\draft
\title{Inflationary Scenarios with Scale-invariant Spectral Tensorial Index}
\author{C\'esar A.\ Terrero-Escalante, Eloy Ay\'on-Beato,
Alberto A. Garc\'{\i}a}
\address{Departamento~de~F\'{\i}sica,
~Centro de Investigaci\'on y de Estudios Avanzados del IPN,
~Apdo.~Postal~14-740,~07000,~M\'exico~D.F.,~M\'exico.}
\maketitle

\begin{abstract}
Next-to-leading order expressions related to Stewart-Lyth inverse
problem are used to determine the inflationary models with a tensorial
power spectrum described by a scale-invariant spectral index. Beyond power-law
inflation,
solutions are characterized by scale-dependent scalar indices. These models 
can be used as assumption on the generation of primordial
perturbations to test for scale dependence of scalar index at
large angular scales. If such a dependence is
detected, a nonzero contribution of gravitational waves to the CMB
spectrum must be expected.
\end{abstract}


\section{Introduction}
\label{sec:Intro}

With measurements carried out by experiments Boomerang and
Maxima-1 \cite{observ} observations of cosmic microwave background
(CMB) anisotropies entered a phase where different theories for
structure formation at our Universe can be falsified. To date,
inflation \cite{Guth,Linde1} is the cosmological model supported
by analysis of these observations \cite{observ2} as favored
theory. Almost any model that produces an accelerated expansion of
the early Universe (i.e., inflation) solves the well-known set of fundamental
problems faced by the Standard Cosmological Model, provided that
the inflationary period lasts long enough (see \cite{LLBook} for
detailed explanations). The simplest inflationary scenario describes the
classical and quantum dynamics of the early Universe dominated by
a single scalar field (inflaton) evolving in a nearly flat
potential.

Analyses such as those in Refs.\cite{observ2} and
\cite{Zalda,Kinney,Covi,Tegmark} consist of maximizing a likelihood
function over the space of model parameters. A given set of
parameters yields a theoretical CMB spectrum to be compared with
observations. The precise number of parameters depends on the
version of the analysis and commonly is as small as 6 and
as large as 11. They are classified as inflationary or
cosmological parameters depending on whether they determine the
initial power spectrum of fluctuations or its posterior evolution.
To set the initial conditions what it is commonly assumed is a
particular kind of single scalar field model namely, power-law
inflation \cite{Lucchin}, characterized by scale-invariant
scalar and tensorial spectral indices differing in unity each from
the other. Amplitudes of tensorial perturbations are often neglected.
The conclusion to be drawn is that, in the corresponding to
observed Universe scales, the actual potential has a strong
similarity with that of power-law inflation. No conclusions can be
made about potential functional form in other scales, particularly
at Planck scales.

When the relative
contribution of kinetic energy to total scalar field energy (known
as $\epsilon$) may be regarded as constant then, to leading order
(LO) in an expansion in terms of $\epsilon$, the scalar and
tensorial indices and $r$, the relative
contribution of primordial gravitational waves to CMB anisotropies, 
are directly related and could be taken
as constants too. In this situation, the model that better
accomplishes the job of fitting the available data is still power-law
inflation, with some distortion taking place for the relation between
spectral indices \cite{HT}. Assumption $\epsilon={\rm const.}$ is
not realistic since the equation of state must change near the
end of inflation in order to return to the Standard Model
expansion rate. Hence, several authors had pointed out that some
scale dependence for the scalar index may be expected
\cite{Mukhanov,Lid4a}. Furthermore, in Ref.~\cite{Covi} it was already found 
currently available data to be compatible with scale-dependent 
scalar index, though this dependence seems to be highly constrained
\cite{Steen}. If a scale-dependent scalar index is the actual case, then 
using a
power-law index would affect the best-fit values of the entire set
of cosmological parameters.

In this work we would like to address the question of which is the
theoretical inflationary potential generating the primordial
fluctuations characterized by a scale-dependent scalar index that
could give the best fit to current as well as to next generation
of observations.

Among other alternatives for deriving the inflaton potential from 
observations (see Ref.~\cite{Lid3a}, and references therein), the Stewart-Lyth
inverse problem (SLIP) was introduced \cite{Nos} as a procedure
consisting of solving a pair of non-linear differential equations
derived in Refs.\cite{SchMi,Benitez} from next-to-leading order (NLO) algebraic
Stewart-Lyth equations for spectral indices \cite{StLy}, and
determining the corresponding inflaton potential. The full power of 
this procedure can be used only when information on the functional forms
of both spectral indices is available. Unfortunately, this information
is rather difficult to be directly obtained from observations and, in addition,
simple functional forms of the spectral indices involve great
difficulties while solving SLIP.
 However, there is an 
alternative way of using the SLIP related expressions. Having 
independent information on the
functional form of one of the indices, the second one can be 
calculated and used as prior for maximizing the likelihood function.
The resulting maximized likelihood will approve or reject the used
spectral indices and the corresponding inflationary potential as
a reliable scenario of the Early Universe.

Next generation of observations is expected to give information 
on CMB
polarization which, in turn, is linked  to $r$ (see Ref.\cite{KK} for 
a review on this subject). Hence, future CMB data, such as MAP
and Planck \cite{MAP} will provide, will determine a range of likely 
values
for tensor-modes parameters, in spite of the subdominance of the
gravitational-waves contribution to CMB anisotropies characteristic
of most inflationary models. Nevertheless, even if a central constant
value of the tensorial parameters can be estimated from these 
observations, it will be almost impossible to detect any scale-
dependence for the tensorial spectral index. With this regard, it 
makes sense to state the problem
as finding out which are the inflationary scenarios, other than
power-law inflation, with scale-invariant spectral tensorial
index. A major theoretical advantage here is that this is one of 
the few cases where the SLIP is analytically solvable.

In this paper we propose two models that could serve to fit the
inflationary perturbations, detecting, if there exist, scale
dependence for the scalar spectral index at large angular scales
while increasing the
scale range and resolution of CMB observations. These models are
obtained in a straightforward manner as SLIP solutions assuming a
constant and almost negligible tensorial index.

In next Sec.\ we briefly describe the theoretical frame for
Stewart-Lyth calculations, the main features of power-law model
upon which these calculations are based and current observations
are fitted, and present Stewart-Lyth algebraic NLO equations for
the spectral indices. In Sec.~\ref{sec:SLeq} SLIP is rewritten
using $\epsilon$ as the basic parameter. Further, we introduce a
criterion that allows us to determine when a given SLIP solution is
consistent with assumptions underlying calculations.
Sec.~\ref{sec:constrain} is devoted to the main aspect of this
manuscript, i.e., how to use NLO expressions related to SLIP to
determine a theoretical potential yielding a spectrum of
perturbations able to fit observations likely to be done in the
near future. We summarize main results obtained in
Sec.\ref{sec:conclu}.

\section{Perturbations produced by inflationary models}
\label{sec:Pert}

Many propositions for inflaton potential can be made fulfilling
the conditions for successful inflation (see Ref.~\cite{LLBook} for
a description of some inflationary models). Hence, the criterion for 
choosing the potential must be the agreement between
theoretical predictions and measurements. Ultimately, testing this
agreement involves the calculation of primordial perturbations
spectra. These spectra are used as initial conditions for the
evolution of perturbations which can be computed through the
transfer functions (using for example the CMBFAST package
\cite{Tegmark}), solutions being compared with current
measurements of CMB anisotropies. The simplest scenario where that
comparison can be carried out is that of a single and real scalar
field rolling down a potential.
In this scenario, a flat Friedmann-Robertson-Walker universe is assumed 
containing a
single scalar field equivalent to a perfect fluid with equations
of motions given by
\begin{eqnarray}
H^2 &=& \frac{\kappa}{3}\left[T + V(\phi)\right],
\label{eq:Friedmann} \\
\ddot{\phi} &+& 3H\dot{\phi} = -V^\prime(\phi) \, ,\label{eq:mass}
\end{eqnarray}
where $\phi$ is the inflaton, $T\equiv\dot{\phi}^{2}/2$, $V(\phi)$ 
the inflationary
potential, $H=\dot{a}/a$ the Hubble parameter, $a$ the scale
factor, dot and prime stand for derivatives with respect to cosmic
time and $\phi$ respectively, $\kappa = 8\pi/m_{\rm Pl}^2$ is the
Einstein constant and $m_{\rm Pl}$ the Planck mass.

In this framework, the first three slow-roll parameters were
respectively defined in Ref.\cite{Lid5} and can be written as
\cite{Lid3a,Nos},
\begin{eqnarray}
\epsilon(\phi) &\equiv& 3T\left[T
+ V(\phi)\right]^{-1} = \frac{2}{\kappa}\left(\frac{H^\prime}{H}\right)^2
= \frac{\kappa T}{H^{2}}\, ,
\label{eq:SRP1} \\
\eta(\phi) &\equiv& -\frac{\ddot{\phi}}{H\dot{\phi}} =
\frac{2}{\kappa}\frac{H^{\prime\prime}}{H} =
\epsilon -\frac{\epsilon^\prime}{\sqrt{2\kappa\epsilon}}
= \kappa \frac{dT}{dH^{2}}\, ,
\label{eq:SRP2} \\
\xi^2(\phi) &\equiv&
\frac{4}{\kappa^2}\frac{H^\prime H^{\prime\prime\prime}}{H^2}
=\epsilon\eta - \left(\frac{2}{\kappa}
\epsilon\right)^{\frac{1}{2}}\eta^\prime
= \, \kappa \epsilon \frac{dT}{dH^{2}}
+2\kappa \epsilon H^{2}\frac{
d^{2}T}{d(H^{2})^{2}}.\label{eq:SRP3}
\end{eqnarray}

Up to a constant, the first slow-roll parameter (\ref{eq:SRP1}) is
a measure of the relative contribution of kinetic energy to total
field energy. By definition $\epsilon\geq0$ and, as it is well
known and we shall see in details further in this manuscript, it
has to be less than unity for inflation to proceed. This last
assertion implies the potential being positive defined. We also
note that while defining second and third slow-roll parameters it
was assumed the potential to be a monotonically decreasing
function. We shall return later to the point of assumptions behind
the definitions and calculations to be used in this manuscript.

Few models of inflation allow exact determination of scalar and
tensorial perturbations. One of them is power-law
\cite{Lucchin}, a scenario of inflation where:
\begin{equation}
\label{eq:PL}
a(t) \propto t^p \, ,
\quad 
H(\phi) \propto \exp\left(-\sqrt{\frac{\kappa}{2p}}\,\phi\right),
\quad
V(\phi) \propto \exp\left(-\sqrt{\frac{2\kappa}{p}}\,\phi\right),
\end{equation}
with $p$ being a positive constant. It follows from Eqs.~(\ref{eq:SRP1}),
(\ref{eq:SRP2}), and (\ref{eq:SRP3}) that in this case the slow-roll
parameters are constant and equal each other, $\epsilon=\eta=\xi=1/p$. Note
that condition $\epsilon<1$ implies $p>1$.

For perturbations produced during power-law inflation 
it can be shown \cite{StLy} that $A_T(k)=10A_S(k)/\sqrt{p}$,
where $A_S(k)$ and $A_T(k)$ stand for normalized scalar and tensorial spectral 
amplitudes,
and $k$ is the wavenumber corresponding to the scale matching the Hubble
radius, $k=aH$. 
Now, through definition of the spectral indices
\begin{equation}
\label{eq:ns}
n_S(k)-1\equiv\frac{d\ln A^2_S(k)}{d\ln k}\, ,
\quad
n_T(k)\equiv\frac{d\ln A^2_T(k)}{d\ln k}
\end{equation}
one can see that for power-law inflation
\begin{equation}
\frac{n_T}{2}=\frac{n_S(k)-1}{2}= \frac{1}{1-p}=-\frac{1}{p}\left(1+\frac{1}{p}+\frac{1}{p^2}
+\cdots\right) \leq 0
\, .
\label{eq:PLns}
\end{equation}
where the inequality follows from above mentioned condition
$\epsilon<1$. Obviously, relation (\ref{eq:PLns}) must be valid to
any order in expansion of $1/(1-p)$. Since spectral indices are
constant, for power-law inflation, definitions (\ref{eq:ns}) can be rewritten 
as
\begin{equation}
\label{eq:PlNs} 
A^2_S(k)=A^2_S(k_0)\left(\frac{k}{k_0}\right)^{n_S-1}
\, ,
\quad
A^2_T(k)=A^2_T(k_0)\left(\frac{k}{k_0}\right)^{n_T}
\, ,
\end{equation}
where $k_0$ is a pivot scale usually taken to be the scale
probed by COBE. From these expressions it is evident that $n_S-1$
and $n_T$ measure the deviation of the spectra amplitudes from
scale-invariance.

Up to the present, there are not general analytic expressions to
calculate power spectra of inflationary models. Based on the power-law 
solution, Stewart and Lyth
\cite{StLy} derived approximated expressions for both spectra
regarding as small the deviation of higher slow-roll parameters
from $\epsilon$ and also deviation of
$\epsilon$ with respect to zero. These
approximations imply the slow-roll parameters to be slowly varying
in time functions. In terms of spectral indices, NLO expressions
are
\begin{eqnarray}
1-n_S(k) &\simeq& 4\epsilon - 2\eta + 8(C+1)\epsilon^2 -
(10C+6)\epsilon\eta + 2C\xi^2, \label{eq:StLythNs} \\
n_T(k) &\simeq& -2\epsilon\left[1+(2C+3)\epsilon-2(C+1)\eta\right],
\label{eq:StLythNt}
\end{eqnarray}
where notation is that of Ref.~\cite{Lid3a} and $C\approx-0.73$ is a
constant. Symbol $\simeq$ is used to indicate that these equations
were obtained using the slow-roll expansion.
Hereafter we shall use equal sign in our calculations, but meaning
of approximation should be added whenever it applies.

For a giving expression of scale factor, the Hubble parameter and
potential are determined, and then substituting definitions
(\ref{eq:SRP1}), (\ref{eq:SRP2}), and (\ref{eq:SRP3}) in
Stewart-Lyth equations (\ref{eq:StLythNs}) and
(\ref{eq:StLythNt}), scale-dependent spectral indices are
obtained. Imposing exact power-law relation $\epsilon=\eta=\xi=1/p$,
expression (\ref{eq:PLns}) is recovered.

Even if it is already possible to include time dependence of
$\epsilon$ at lowest order, we shall show here that some
interesting cases able to answer the question stated in this paper 
are
only considered if NLO
expressions are used. NLO expressions for the spectra have been
tested and found to provide a high accuracy for theoretical
perturbations calculations \cite{Lid4}. Even more, some authors
have stressed that NLO expressions in terms of $\epsilon$ will be
compulsory in order to match analytic results with data to be
obtained in the near future \cite{Lid3a,Schwarz}.

For Eq.~(\ref{eq:StLythNt}) constrain $n_T\leq0$ remains
valid, even for scale-dependent tensorial index. Assume $n_T>0$
then, from Eq.~(\ref{eq:StLythNt}), we obtain inequality
$1\leq1+\epsilon < 2(C+1)(\eta-\epsilon)$. Evaluating $C=-0.73$
yields $\eta-\epsilon \geq 1.852$, in contradiction with the
approximation $\left|\eta-\epsilon\right|<1$ used to derive
Eq.~(\ref{eq:StLythNt}). Thus, in general,
\begin{equation}
n_T(k)\leq 0 \, . \label{eq:condnT}
\end{equation}
For the kind of models we are considering here, even if tensorial
perturbations are large enough to have a detectable imprint in CMB
anisotropies, according with definition (\ref{eq:ns}), their
amplitudes will vanish in wave longitudes to be probed by most
gravity waves interferometers \cite{LLBook}.
A similar analysis for Eq.~(\ref{eq:StLythNs}) yields that,
although models with $n_S<1$ are favored,
not particular constrain exist upon values of $n_S$, thus allowing
models with so called, blue spectra, i.e., $n_S>1$.

\section{Stewart-Lyth Inverse Problem}
\label{sec:SLeq}

In Ref.\cite{Nos} it was shown that using definitions (\ref{eq:SRP1}),
(\ref{eq:SRP2}), (\ref{eq:SRP3}) and defining $\tau \equiv \ln \,H^{2}$, 
$\delta (k)\equiv n_{T}(k)/2$
and $\Delta (k)\equiv [n_{s}(k)-1]/2$, the indices equations in terms of 
the first
slow-roll parameter $\epsilon $ and its derivatives with respect
to $\tau$ ($\hat{\epsilon}\equiv d\epsilon/d\tau $ and
$\hat{\hat{\epsilon}}\equiv d^{2}\epsilon /d\tau ^{2}$), in a
straightforward manner, can be written as \cite{SchMi,Benitez}
\begin{eqnarray}
2C\epsilon \hat{\hat{\epsilon}}-(2C+3)\epsilon \hat{\epsilon}-\hat{\epsilon}
+\epsilon ^{2}+\epsilon +\Delta &=&0\,, \label{eq:MSch1} \\
2(C+1)\epsilon \hat{\epsilon}-\epsilon ^{2}-\epsilon -\delta &=&0\,.
\label{eq:MSch2}
\end{eqnarray}

Given expressions for scale-dependent spectral indices,
corresponding inflaton potential can be found by solving
Eqs.~(\ref{eq:MSch1}) and (\ref{eq:MSch2}) for $\epsilon$, and
using definitions of first slow-roll parameter (\ref{eq:SRP1}).
This procedure is what we called Stewart-Lyth inverse problem \cite{Nos}.

When SLIP was introduced in Ref.~\cite{Nos}, the inflaton
potential corresponding to a given solution of differential
equations  (\ref{eq:MSch1}) and (\ref{eq:MSch2}), was determined
as a parametric function of $\tau$. Solutions of these equations
are expressed with $\tau$ as an explicit function of $\epsilon$
and are difficult to convert to expressions for $\epsilon$ as
explicit functions of $\tau$. Hence, it seems reasonable to look
for a similar procedure, in terms of the first slow-roll
parameter. Moreover, as we shall see later, using $\epsilon$ as a
parameter allows us to analyze the solution restricted to the
interval of $\phi$ where inflation is feasible, i.e.,
$0\leq\epsilon<1$.

The expression for the potential as a function of $\epsilon$ remains the same
that in Ref.~\cite{Nos} and is obtained from definition (\ref{eq:SRP1}),
\begin{equation}
V(\epsilon)= \frac{1}{\kappa}\left(3-\epsilon\right)
\exp\left[\tau(\epsilon)\right]
\,,
\label{eq:PotentialE}
\end{equation}
but here, instead substituting the first slow-roll parameter as function of
$\tau$, we substitute $\exp[\tau(\epsilon)]$.
On the other hand, using Eq.~(\ref{eq:SRP1}) \cite{Nos},
\begin{equation}
\phi(\tau)=-\frac{1}{\sqrt{2\kappa}}
\int\frac{d\tau}{\sqrt{\epsilon(\tau)}}
+ \phi_0\, ,
\label{eq:PhiT}
\end{equation}
where $\phi_0$ is an integration constant. Changing variables,
and substituting $\hat{\epsilon}$ from the first order equation (\ref{eq:MSch2}),
\begin{equation}
\phi(\epsilon)= -\frac{2(C+1)}{\sqrt{2\kappa}}
\int\frac{\sqrt{\epsilon}d\epsilon}{\epsilon^2+\epsilon+\delta}
+ \phi_0\, .
\label{eq:Phi}
\end{equation}

This way, the inflaton potential is given by the parametric function,
\begin{equation}
V(\phi)= \left\{
\begin{array}{c}
\phi(\epsilon)\, ,  \\
V(\epsilon)\, .
\end{array}
\right.
\label{eq:FVphi}
\end{equation}

Here is very important to recall that for SLIP solution
(\ref{eq:FVphi}) to be unique, $\epsilon(\tau)$ must be solution
of both equations (\ref{eq:MSch1}) and (\ref{eq:MSch2}). Need of
information on the tensorial modes is also a conclusion stressed
by Lidsey et al. \cite{Lid3a} in their report about perturbative
reconstruction of inflaton potential. With this regards, one can
see that solution $\epsilon=1/p$ to Eqs. (\ref{eq:MSch1})
and (\ref{eq:MSch2}) automatically implies
$\Delta=\delta=1/(1-p)$ in full correspondence with relation (\ref{eq:PLns}).

\subsection{Consistency criterion for SLIP solutions}
\label{subsec:propsol}

Solving SLIP is not enough to state that solutions have any
physical meaning. We recall that our calculations are
based on several assumptions regarding the form of the potential
(behavior of the inflaton as function of cosmic time) and range of
slow-roll parameters. Hence, for SLIP to yield consistent results,
conditions arising from this assumptions should be fulfilled by
the obtained inflaton and its potential. Let us analyze these
conditions in detail.

To derive Eqs.~(\ref{eq:StLythNs}) and (\ref{eq:StLythNt}) and
SLIP solution (\ref{eq:FVphi}),
no particular assumption was made about initial conditions for $\phi$
nor for its expected value thus, solutions of SLIP are not
constrained to chaotic or new inflation nor to a particular energy
scale. Furthermore, no assumption was neither made about the
potential convexity so, in principle, it could be in any of
categories related to classification given in Ref.~\cite{Kinney}
and moreover, the same potential could have features
characteristic of different categories. This is a consequence of
dealing with NLO expressions, i.e., allowing a larger variation of
slow-roll parameters during inflation. Particularly, SLIP
solutions can be associated with a hybrid scenario of inflation
\cite{Linde2}. Next, we shall analyze possible constrains upon
solutions of SLIP. Deriving Eq.~(\ref{eq:Friedmann}) with respect to
cosmic time and inserting Eq.~(\ref{eq:mass}) it is obtained,
\begin{equation}
T = - \frac{1}{\kappa}\dot{H}\, .
\label{eq:TvsHdot}
\end{equation}
Now, taking into account that $H^\prime=\dot{H}/\dot{\phi}$,
to determine a sign for $H^\prime$ it is necessary to fix the
inflaton behavior as a function of cosmic time. In this paper, it was assumed
that $\dot{\phi}>0$, and, correspondingly, $H^\prime<0$. Further,
comparing  definitions in Eq.~(\ref{eq:SRP1}) for
the first slow-roll parameter and eliminating $\epsilon$ yields,
\begin{equation}
T = \frac{2}{\kappa^2}{H^\prime}^2\, ,
\label{eq:TvsHpr}
\end{equation}
and after substituting Eq.~(\ref{eq:TvsHpr}) in the equation of motion 
(\ref{eq:Friedmann})
and deriving with respect to $\phi$ we obtain,
\begin{equation}
V^\prime = \frac{2}{\kappa}
\left(3-\eta\right)HH^\prime
\, ,
\label{eq:VprvsHpr1}
\end{equation}
where the definition of the second slow-roll parameter (\ref{eq:SRP2}) 
was used.

Eqs.~(\ref{eq:StLythNs}) and (\ref{eq:StLythNt}) were obtained
using the slow-roll expansion, i.e, the absolute
value of $\eta$ should be close to the value of $\epsilon$ which,
in turn, should be near zero. Therefore, $(3-\eta)>0$ and the sign
of $V^\prime$ is determined by that of $H^\prime$. Hence,
according to the previous assumptions, the potential must be a
monotonically decreasing function of the inflaton. Summarizing,
any feasible solution of SLIP should fulfill the following
conditions:
\begin{equation}
\left\{
\begin{array}{rcl}
\dot{\phi}&>&0\, , \\
V^\prime(\phi)&<&0\, .
\end{array}
\right.
\label{eq:phiConds}
\end{equation}
These correspond to conditions for the inflaton to roll down the
potential from lower to higher values of the scalar field.
At least in the case we shall analyze here, it seems to be useful
to write conditions (\ref{eq:phiConds}) in an equivalent manner.
Note that $d\tau/dt=2\dot{H}/H<0$, then conditions
(\ref{eq:phiConds}) imply
\begin{equation}
\left\{
\begin{array}{rcl}
\hat{\phi}&<&0\, ,   \\
\hat{V}&>&0\, ,
\end{array}
\right.
\qquad
\Longleftrightarrow
\qquad
\left\{
\begin{array}{rcl}
\hat{\epsilon}\frac{d\phi}{d\epsilon}&<&0\, ,  \\
\hat{\epsilon}\frac{dV}{d\epsilon}&>&0\, .
\end{array}
\right.
\label{eq:epsConds}
\end{equation}
Similar conditions could be derived for the case of an inflaton rolling down
from higher to lower values by properly choosing the behavior of
the inflaton as a function of cosmic time. For that case conditions 
(\ref{eq:epsConds}) will read,
\begin{equation}
\left\{
\begin{array}{rcl}
\hat{\epsilon}\frac{d\phi}{d\epsilon}&>&0\, ,  \\
\hat{\epsilon}\frac{dV}{d\epsilon}&>&0\, .
\end{array}
\right.
\label{eq:mepsConds}
\end{equation}
In fact, of all the expressions used here, the sign of $\dot{\phi}$
only affects Eq.~(\ref{eq:Phi}), the modification being
$\phi(\epsilon)\rightarrow-\phi(\epsilon)$. That changes the
SLIP solutions by the mirror equivalent solutions.

On the other hand, inflation is defined as a period where scale
factor grows accelerately, i.e, $\ddot{a}>0$. This is equivalent
to say that $d(H^{-1}/a)/dt<0$, definition remarking that,
during inflation, the comoving Hubble radius decrease with time.
 Deriving, using expression (\ref{eq:TvsHdot}) and
definition (\ref{eq:SRP1}) we obtain already mentioned upper
value for first slow-roll parameter, i.e., $\epsilon<1$. By
definition $\epsilon\geq0$, then criteria (\ref{eq:epsConds}) and
(\ref{eq:mepsConds}) must be tested in the interval $\epsilon\in[0,1)$.
Looking at graphs of inflaton and its potential as functions of
$\epsilon$, and plots of $\epsilon(\tau)$ in the corresponding
range, one is able to find out whether SLIP solutions will be
consistent with underlying assumptions. Note that these conditions
are sufficient for given potential to be inflationary but they do
not ensure the inflationary epoch to be long enough.

\section{Models with constant tensorial spectral index}
\label{sec:constrain}

To settle down the cosmological parameters, the set containing the 
parameters that determine
the realization of the cosmological model together with the parameters that
determine the initial conditions is tuned in order to maximize a
likelihood function \cite{observ2,Zalda,Kinney,Covi,Tegmark}. Commonly, the 
initial power spectra are
set to the form corresponding to power-law inflation with
negligible amplitudes for primordial gravitational waves. The
assumption behind is that during the inflationary lapse where
quantum fluctuations were imprinted in scales currently reentering
our causal Universe, slow-roll parameters behave roughly as
constants.

An important parameter is the tensor-scalar ratio of contribution
to the CMB spectrum which can be defined as (see Ref.~\cite{LLBook}
for alternative definitions),
\begin{equation}
r \equiv 12.4\frac{A_T^2}{A_S^2}\, .
\label{eq:r}
\end{equation}
In principle, the value of $r$ can be estimated
from observations of CMB polarization \cite{KK}. To NLO, $r$ is related
with the tensorial index by \cite{Lid3a}
\begin{equation}
n_T \simeq -2\frac{A_T^2}{A_S^2}(1+3\epsilon-2\eta)\, .
\label{eq:rnT}
\end{equation}
From Eqs.~(\ref{eq:StLythNs}) and (\ref{eq:StLythNt}), LO
expressions are recovered by neglecting second order terms of
slow-roll parameters, and from Eq.~(\ref{eq:rnT}) by setting the
expression within parenthesis equal to unity. This way, if in the
desired scale range $\epsilon$ and $\eta$ can be approximated by
constant values, to LO, $n_S$, $n_T$ and $r$ must be regarded
as constants too. Hence, if some information on $r$ is available,
power-law inflation can still be the model providing the best fit
to data given some distortion of the relation between
indices. 
If some
degree of scale dependence is hidden in the CMB, 
the error of
assuming power-law will be reflected in the best-fit
values of remaining parameters. Hence, to consider $\epsilon\simeq
\rm const.$ is a big restriction regarded as fair if the fit
of data using a constant scalar index give a good result.
Nevertheless, even now, such a good overall fit can also be achieved 
for
some potentials yielding scale dependent scalar index, for
example, running mass models \cite{Covi} though the recent results
in Ref.~\cite{Steen} constrain this dependence to be very weak. Several
authors have pointed out that observations with higher resolution
and wider scale-range to be provided by satellites Planck and Map,
and galaxies surveys should be able to discern a time dependent
$\epsilon$ \cite{Lid4}, making of power-law a poor assumption for
the inflationary period. Therefore, question arises of which model
could provide the best fit to upcoming data. To answer this
question, an option is to find out if there exist models with
slowly-varying tensorial and scalar spectral indices which, using
current data, can be accurately fitted by a power-law and smoothly
departs from power-law while making broader the range of scales or
increasing the resolution of measurements. 
With this aim,
one can make assumptions on the functional form for $n_T$ and look
for solutions of SLIP and corresponding functional forms of the
scalar spectral index which, in turn, can be compared with
observations of CMB anisotropies. In general, the tensorial spectral index 
should be a scale-dependent function slowly varying close to zero
thus, a well-based assumption for $\delta$ is
\begin{equation}
\delta(\ln k)= \delta(\ln k_0) + a_1\ln\frac{k}{k_0} + a_2\ln^2\frac{k}{k_0}
+ \cdots
\, .
\label{eq:dexp}
\end{equation}
The first reliable observation of $r$ to be obtained is hardly
expected to detect any dependence on the scale \cite{KK}. Hence, from this
information
only a constant value for $n_T$ would be estimated. With this regard, in this
paper, the analysis will be
restricted to zero order in expansion (\ref{eq:dexp}). This approach has also
the advantage of dealing with one of the few cases when SLIP is analytically
solvable. We remark that to LO
it is a nonsense to consider a
constant tensorial index while regarding a time dependent
$\epsilon$. Solutions presented here can be
obtained only by using NLO expressions related to SLIP.

\subsection{Constant scalar and tensorial indices}

Let us start by proving that using $\delta=C_1$ and $\Delta=C_2$ as SLIP
input (with $C_1$ and $C_2$ being some constants) just yields
power-law solution. Recall that power-law inflation has a 
number of
characteristic features: slow-roll parameters are constant and equal
each other, spectral indices are constant, and spectra given by these
indices are red tilted in the same magnitude from Harrison-Zeldovich
spectra, i.e., $C_1=C_2\neq 0$. To proceed with, we note that
in Eqs.~(\ref{eq:MSch1}) and (\ref{eq:MSch2}) $\epsilon$ and its
derivatives depend on $\tau$ while $\Delta$ and $\delta$
explicitly depend on $k$. So, we shall rewrite these equations in
terms of $k$ in the same fashion it was done in
Ref.~\cite{NosIII}. 
 The conversion between the inflaton values and wavenumbers while
crossing the Hubble
radius can be done using expression \cite{Lid3a}
\begin{equation}
\frac{d\ln k}{d\phi}=\frac{\kappa}{2}\frac{H}{H^\prime}\left(\epsilon-1\right).
\label{eq:k2phi}
\end{equation}
Using Eq.~(\ref{eq:k2phi}) it is obtained,
\begin{equation}
\frac{d\ln k}{d\tau}=\frac{1}{2}\frac{\epsilon-1}{\epsilon}\, .
\label{eq:tau2k}
\end{equation}
After conversion to derivatives in term of $\ln k$, Eqs.~(\ref{eq:MSch1})
and (\ref{eq:MSch2}) become \cite{NosIII},
\begin{eqnarray}
\frac{C(\epsilon-1)^2}{2\epsilon}\tilde{\tilde{\epsilon}}
+\frac{C(\epsilon-1)}{2\epsilon^2}\tilde{\epsilon}^2
-\left[(2C+3)\epsilon+1\right]\frac{\epsilon-1}{2\epsilon}\tilde{\epsilon}
+\epsilon^2+\epsilon+\Delta&=&0,\label{eq:Eqk1}\\
(C+1)(\epsilon-1)\tilde{\epsilon}-\epsilon^2-\epsilon-\delta&=&0,
\label{eq:Eqk2}
\end{eqnarray}
where $\tilde{\epsilon} \equiv d\epsilon/d\ln k$ and $\tilde{\tilde{\epsilon}}
\equiv d^2\epsilon/d(\ln k)^2$.

Differentiating Eq.~(\ref{eq:Eqk2}) with respect to $\ln k$ we can
replace expressions for $\tilde{\epsilon}$ and
$\tilde{\tilde{\epsilon}}$ obtained from this equation into
Eq.~(\ref{eq:Eqk1}) and the following algebraic expression for
$\epsilon$ is obtained:
\begin{equation}
\label{eq:Eq4degree}
\epsilon(k)^4+P\epsilon(k)^3+Q(\tilde{\delta},\delta, \Delta)\epsilon(k)^2
+R(\tilde{\delta},\delta)\epsilon(k)+S(\delta)=0,
\end{equation}
where
\begin{eqnarray}
\nonumber
P&=&C+2,\\
\nonumber
Q(\tilde{\delta},\delta, \Delta)&=&-(C+1)\left [C\tilde{\delta}-(2C+3)\delta
+2(C+1)\Delta-1 \right ],\\
\nonumber
R(\tilde{\delta},\delta)&=&(C+1)C\tilde{\delta}+(2C+1)\delta,\\
S(\delta)&=&C\delta^2.
\nonumber
\end{eqnarray}
Roots of Eq.~(\ref{eq:Eq4degree}) can be calculated but they are
very complicated expressions not necessary for further analysis.
They can be simply written as
\begin{equation}
\label{eq:Roots4}
\epsilon_i=\epsilon_i(\Delta, \delta, \tilde{\delta}),
\end{equation}
where $i=1,\dots,4$. Thus, for $\delta=C_1$ and $\Delta=C_2$ the
solution is $\epsilon=\rm const.$. Substituting back this solution
into Eqs.~(\ref{eq:Eqk1}) and (\ref{eq:Eqk2}) we finally obtain
$\delta=\Delta$. Thus, as it was expected, power-law 
inflation is a trivial solution
of our problem. Notice that, according with Eq.~(\ref{eq:PL}) any inflationary 
potential that behaves like a
decaying exponential for a range of $\phi$ corresponding to scales
currently probed can be, in this range, approximated by a power-law. While this
will offer good enough information on the inflationary period corresponding to
currently observed Universe, it perhaps will hide clues about physics on
higher energy scales.

\subsection{Scale-dependent scalar index and constant tensorial index}
\label{subsec:Denk}

Previously it has been stressed the relevance of Eq.~(\ref{eq:MSch2}) for
constraining solutions of Eq.~(\ref{eq:MSch1}). As was already proved,
for constant $\delta$, a trivial SLIP solution is power-law
inflation, corresponding to $\delta=\Delta$. We shall analyze here
remaining solutions involving scale-dependent $\Delta$.
The methodology used in this research to solve the problem of finding models
with scale-invariant tensorial index but scale-dependent scalar index
consists of solving Eq.~(\ref{eq:MSch2}) for $\epsilon$ using as input 
a constant $\delta$, substituting the
solutions on Eq.~(\ref{eq:MSch1}) and deriving the corresponding functional
forms for the inflaton potential and $\Delta$.

With the aim of obtaining an expression
for the scalar spectral index as function of scale, let us first 
to express $\Delta$ as a function of
$\epsilon$. It can be derived from algebraic
Eq.~(\ref{eq:Eq4degree}) considering a constant $\delta$,
\begin{equation}
\Delta(\epsilon)=\frac{1}{2\left(C+1\right)^2}\left\{ \epsilon^2
+\left(C+2\right)\epsilon
+\left(C+1\right)\left[\left(2C+3\right)\delta+1\right]
+\left(2C+1\right)\frac{\delta}{\epsilon}
+C\frac{\delta^2}{\epsilon^2} \right\} \, .
\label{eq:De}
\end{equation}
Now, we shall look for an expression of the comoving scale as a function 
of $\epsilon$. This relation is obtained by integrating
Eq.~(\ref{eq:Eqk2}):
\begin{equation}
\ln k(\epsilon)=\frac{(C+1)}{2}\left(\ln \left|\epsilon^2+\epsilon
+\delta\right| -
3\int\frac{d\epsilon}{\epsilon^2+\epsilon+\delta}\right)
\, .
\label{eq:ke}
\end{equation}
This way, the scalar spectral index can be expressed as a parametric function 
given by
\begin{equation}
n_S(k)= \left\{
\begin{array}{cc}
k(\epsilon)\, , &  \\
2\Delta(\epsilon)+1\, . &
\end{array}
\right.
\label{eq:Dk}
\end{equation}

The relevant inflationary parameters are the scalar and 
tensorial spectra amplitudes evolved through the transfer functions into the
CMB anisotropies spectrum.
Having Eq.~(\ref{eq:Dk}), a parametric expression for the scalar amplitudes 
$A_S(k)$ can be
readily derived using definition (\ref{eq:ns}) and Eq.(\ref{eq:Eqk2}):
\begin{equation}
A_S(k)= \left\{
\begin{array}{cc}
k(\epsilon)\, , &  \\
A_S(\epsilon)\, , &
\end{array}
\right.
\label{eq:Ask}
\end{equation}
with $A_S(\epsilon)$ given by
\begin{equation}
A_S(\epsilon) = A_0\exp\left[(C+1)\int{ \Delta(\epsilon)
\frac{\epsilon-1}{\epsilon^2+\epsilon+\delta}
d\epsilon}\right]
\, ,
\label{eq:Ase}
\end{equation}
where $A_0$ is an integration constant, $\Delta(\epsilon)$ is given by
expression (\ref{eq:De}) and the scale as function of $\epsilon$ by 
Eq.~(\ref{eq:ke}). 

As it was discussed in Sec.~\ref{sec:Pert} while
analyzing constrain given by Eq.~(\ref{eq:condnT}), we have that possible 
values of the tensorial index are restricted to the interval $\delta\leq 0$. 
With this regard, in Secs.~\ref{subsec:sold} and
\ref{subsec:general} we shall solve the problem of finding models with
scale-invariant tensorial index and scale-dependent scalar index for 
$\delta=0$ and $\delta<0$ respectively.

\subsubsection{Null tensorial index, $\delta=0$}
\label{subsec:sold}

We begin by analyzing the case $\delta=0$ which, being easy to explicitly 
solve, is a good example of using SLIP
to test the reliability of given functional form for tensorial
index and of corresponding analysis of solutions. However, we shall see
that the results fail to match current observations.

\begin{figure}[t]
\centerline{\psfig{file=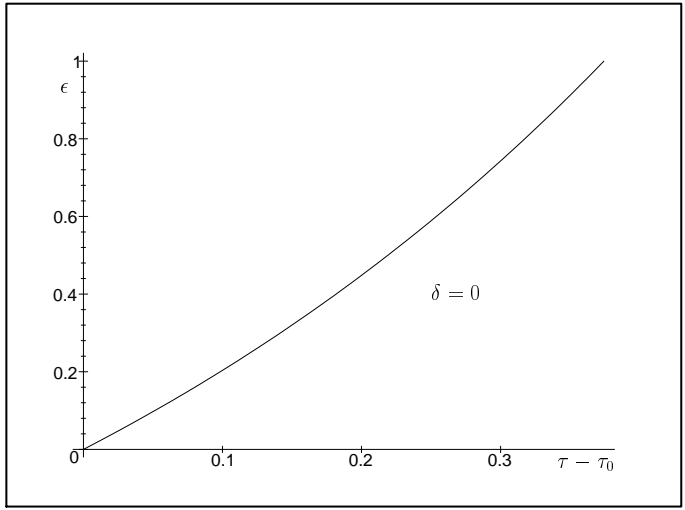,width=9.5cm}}
\caption{Solution of the first order equation for $\delta=0$.}
\label{fig:d0et}
\end{figure}

Explicit integration of Eq.~(\ref{eq:MSch2}) with $\delta=0$ yields,
\begin{equation}
\exp(\tau-\tau_0)=\left(\epsilon+1\right)^{2(C+1)}\, ,
\label{eq:FO4}
\end{equation}
where $\tau_0$ is the integration constant. This solution, first reported in
Ref.~\cite{Nos}, is plotted in Fig.~\ref{fig:d0et}.
Corresponding to $\delta=0$ expression for $\phi(\epsilon)$
is obtained by straightforward integration of Eq.~(\ref{eq:Phi}),
\begin{equation}
\phi(\epsilon)= -\frac{4(C+1)}{\sqrt{2\kappa}}
\arctan\left(\sqrt{\epsilon}\right)
+\phi_0\, .
\label{eq:pedo}
\end{equation}
Expression for $V(\epsilon)$ is obtained by substituting solution
(\ref{eq:FO4}) in Eq.~(\ref{eq:PotentialE}),
\begin{equation}
V(\epsilon)=V_0(3-\epsilon)(\epsilon+1)^{2(C+1)} \, ,
\label{eq:Sol5}
\end{equation}
where $V_0=\exp(\tau_0)/\kappa$. Separating $\epsilon$ in the
expression for $\phi$ and substituting in Eq.~(\ref{eq:Sol5}) finally 
yields for the inflaton potential,
\begin{figure}[h]
\centerline{\psfig{file=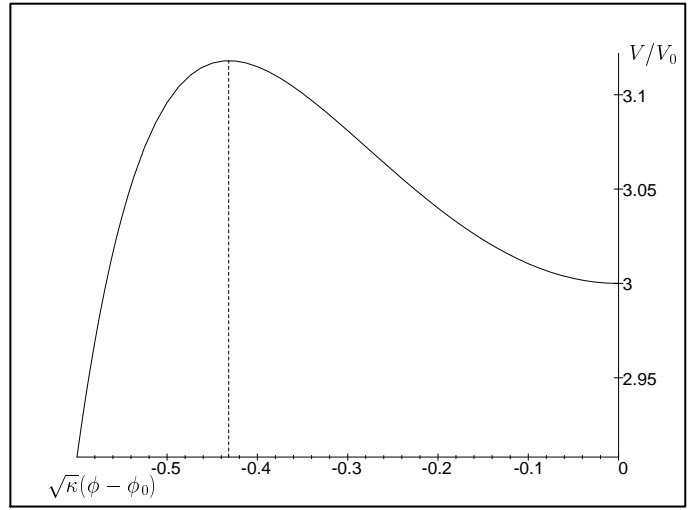,width=9.5cm}}
\caption{SLIP solution for $\delta=0$.}
\label{fig:d0Vph}
\end{figure}
\begin{equation}
V(\phi)=V_0\frac{3-\tan^2\left[\frac{\sqrt{2\kappa}}
{4(C+1)}(\phi-\phi_0)\right]}
{\cos^{4(C+1)}\left[\frac{\sqrt{2\kappa}}
{4(C+1)}(\phi-\phi_0)\right]}\, .
\label{eq:Vdelta0}
\end{equation}
Plot of this solution for range corresponding to
$0\leq\epsilon<1$ is presented in Fig.~\ref{fig:d0Vph}.

Looking at expression (\ref{eq:Vdelta0}) and
Fig.~\ref{fig:d0Vph}, one can conclude that exist an interval of $\phi$ where
the potential is not consistent with the assumption $\dot{\phi}>0$,
i.e., where the potential increases with the
inflaton value. 
One can wonder whether there is another 
sector of this potential where inflation can take place. Indeed, such a sector
can exist but the primordial fluctuations generated during the corresponding
inflaton rolling down could be different to those assumed as SLIP input.
Note that the range of $\phi$ used to plot solution
(\ref{eq:Vdelta0}) was possible to determine using
Eq.~(\ref{eq:pedo}) which describe the inflaton as a function of the first
slow-roll parameter. The valid range of values for $\epsilon$ determine the
corresponding inflaton values. 

\begin{figure}[h]
\centerline{\psfig{file=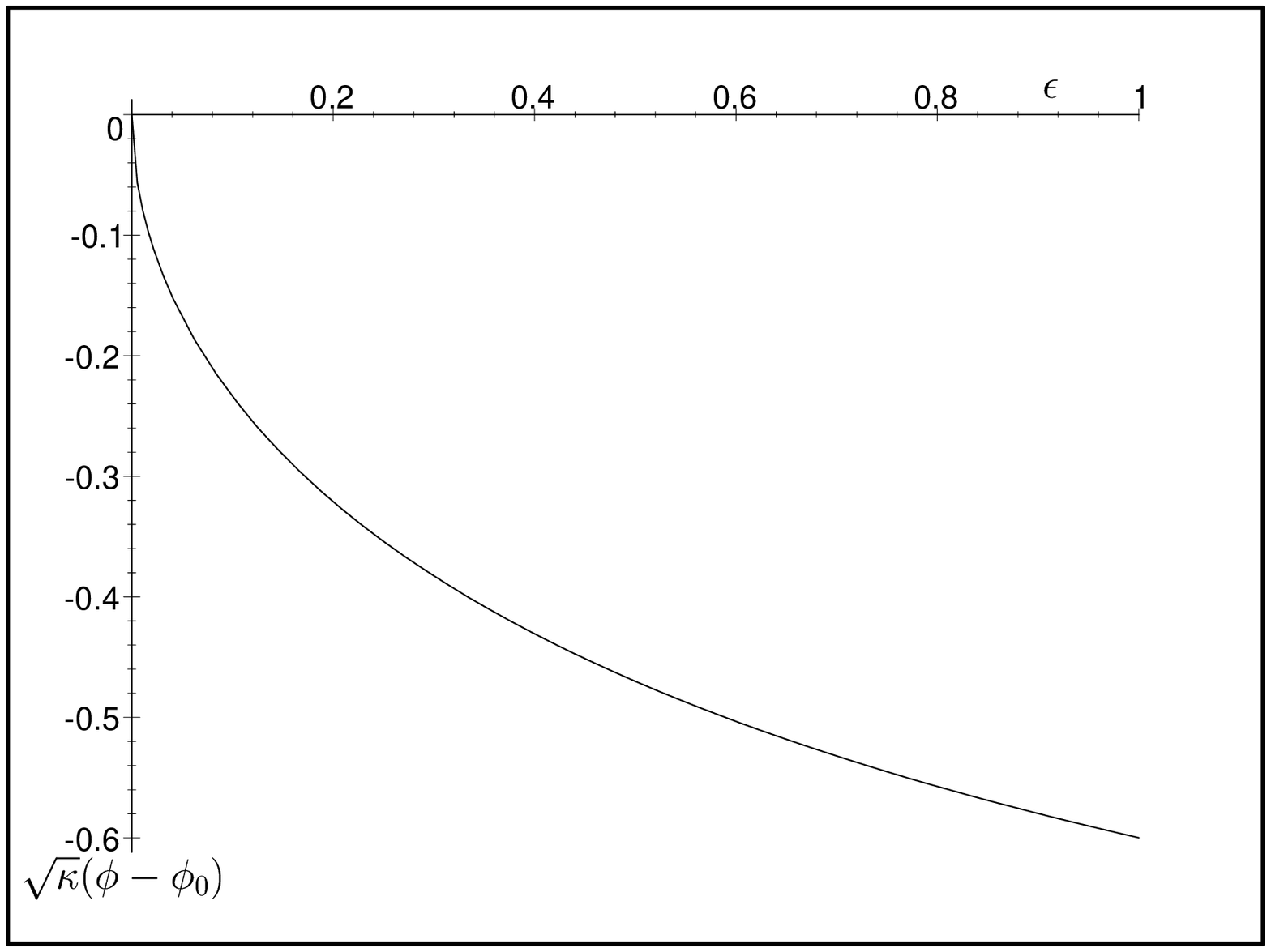,width=9.5cm}
\psfig{file=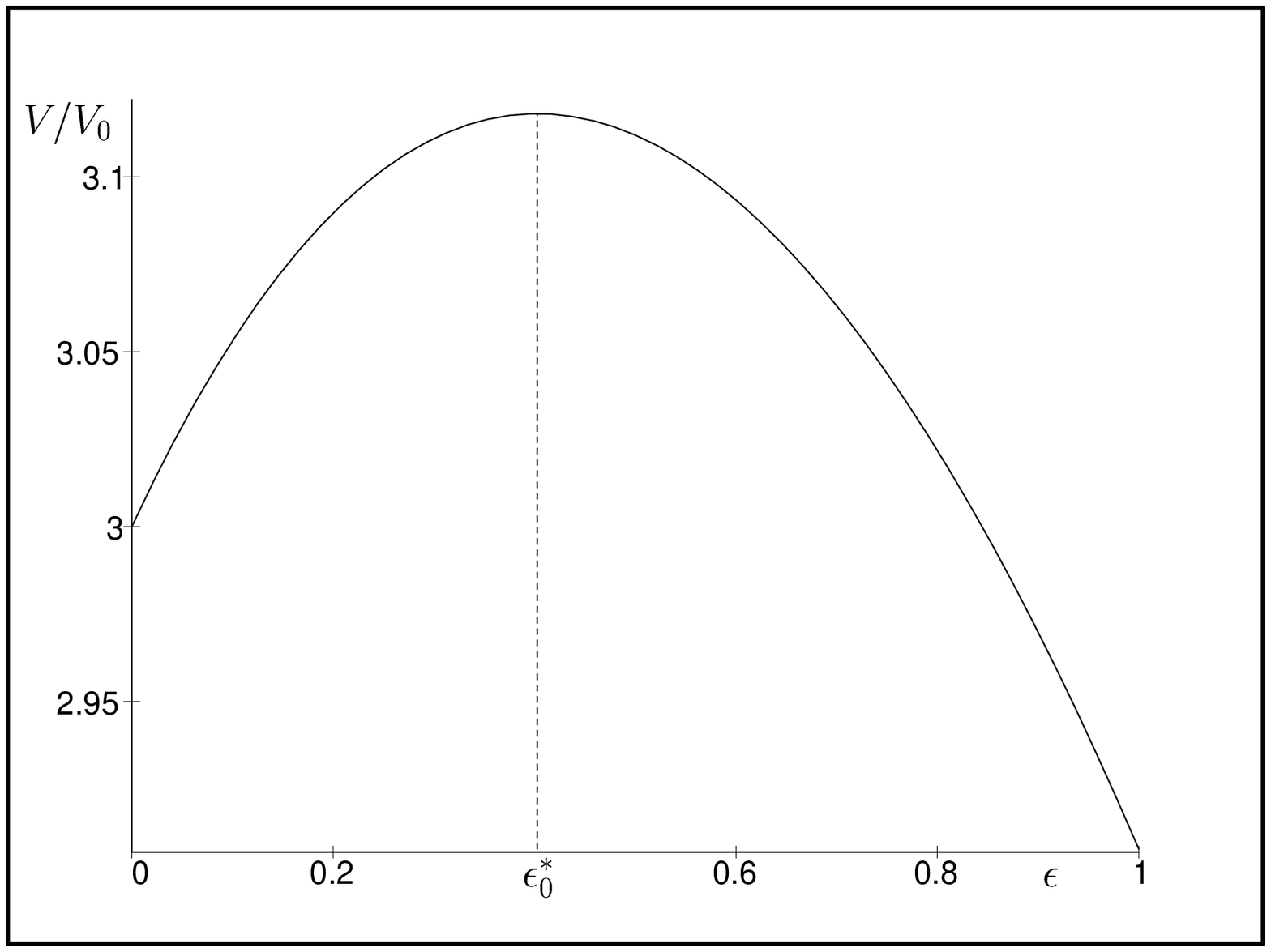,width=9.5cm}}
\caption{Inflaton and its potential as functions of the first
slow-roll parameter for $\delta=0$.}
\label{fig:d0pVe}
\end{figure}

\begin{figure}[h]
\centerline{\psfig{file=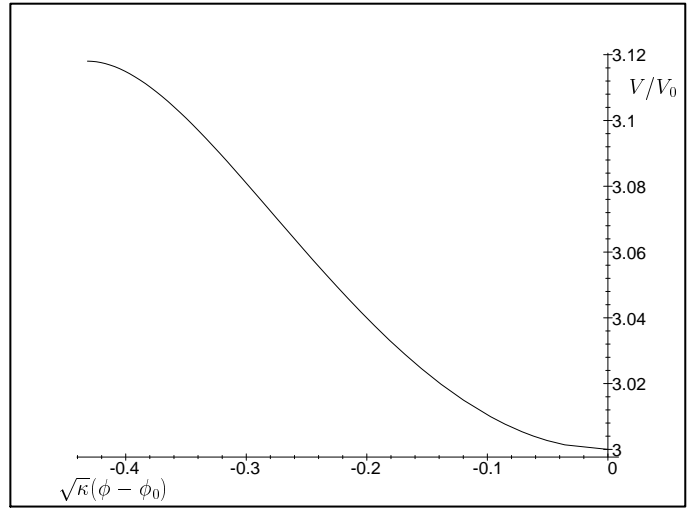,width=9.5cm}}
\caption{Consistent SLIP solution for $\delta=0$.}
\label{fig:d0Vp1}
\end{figure}

Let us look at plots of $\phi$ and $V$ as
functions of $\epsilon$, Fig.~\ref{fig:d0pVe}.
Regarding criterion (\ref{eq:epsConds}) and Figs.~\ref{fig:d0et},
and \ref{fig:d0pVe}, it can be concluded that solution is
consistent only for $\epsilon\in[0,\epsilon_0^*]$ with
$\epsilon_0^*\equiv (6C+5)/(2C+3)\simeq 0.4$. This way, the
correct functional form for the potential with $\delta=0$ is that
in Fig.~\ref{fig:d0Vp1}.
Observe that for this case the potential curvature undergoes
changes. Such a behavior for the potential is obtained as a SLIP solution 
due to the use of NLO expressions.

After integrating Eq.~(\ref{eq:Eqk2}), the comoving number $k$ as
a function of $\epsilon$ is given by
\begin{equation}
k=k_0\left(\frac{(\epsilon+1)^2}{\epsilon}\right)^{C+1}\, ,
\label{eq:k4}
\end{equation}
with $k_0$ being the integration constant. From here, $\epsilon(k)$ is
obtained,
\begin{equation}
\epsilon(k)=\frac12\left\{\left(\frac k{k_0}\right)^{1/\left(C+1\right)} -2
-\sqrt{ \left(\frac k{k_0}\right)^{1/\left(C+1\right)}
\left[\left(\frac k{k_0}\right)^{1/\left(C+1\right)}-4\right] }
\right\} \, ,
\label{eq:ek}
\end{equation}
where the root was chosen corresponding to $0\leq\epsilon<1$.

\begin{figure}[h]
\centerline{\psfig{file=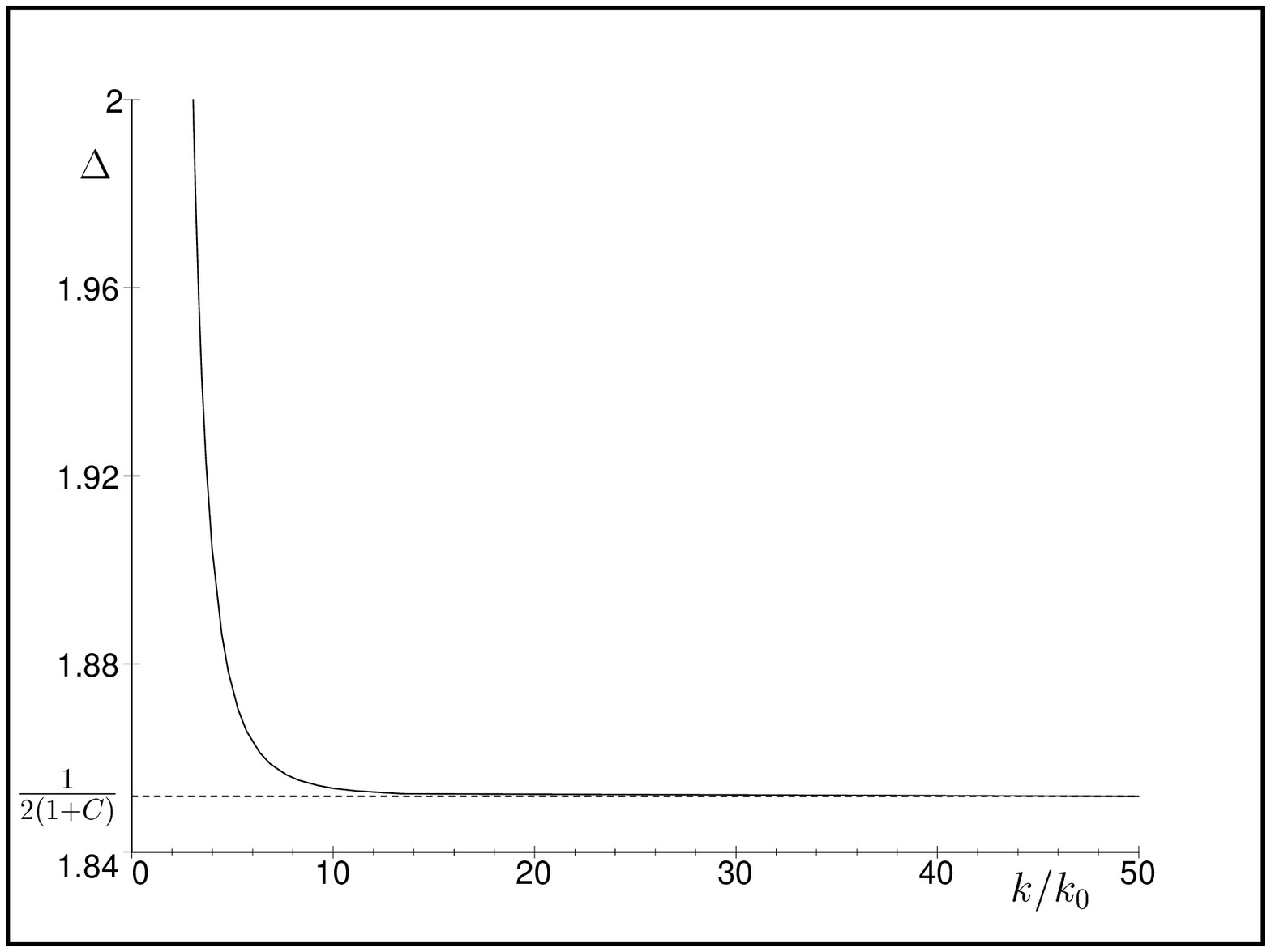,width=9.5cm}
\psfig{file=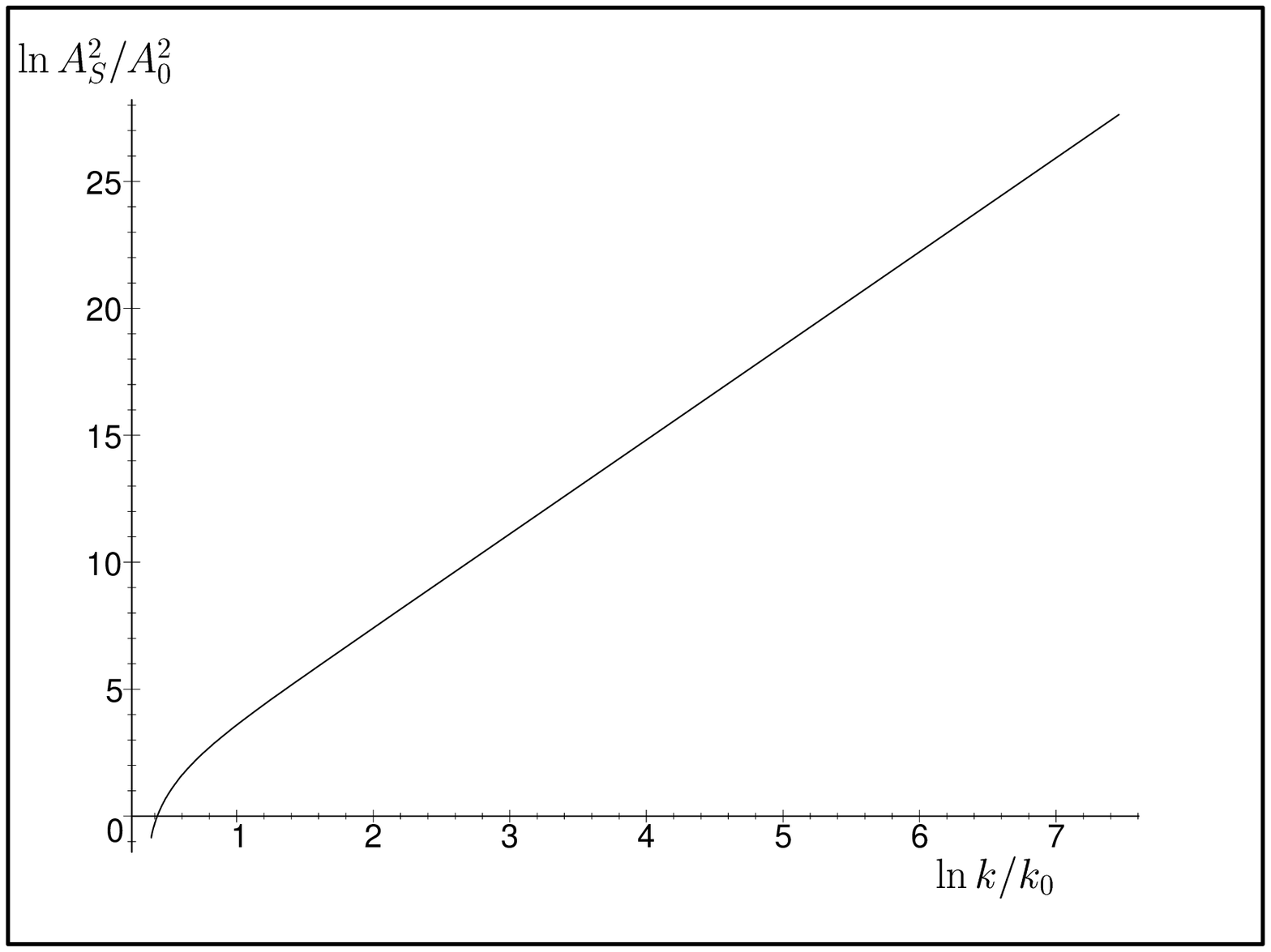,width=9.5cm}}
\caption{$\Delta$ as a function of the comoving number $k$
and $\ln A^2_S$ as a function of the $\ln k$ for $\delta=0$.}
\label{fig:d0Dk}
\end{figure}

Substituting expression (\ref{eq:ek}) in Eq.~(\ref{eq:De}) with $\delta=0$,
the corresponding expression for $\Delta(k)$ is,
\begin{eqnarray}
\Delta(k) &=& \frac1{8\left(C+1\right)^2}\left\{
\left(\frac k{k_0}\right)^{1/\left(C+1\right)}
-\sqrt{\left( \frac k{k_0}\right) ^{1/\left(
C+1\right) }\left[ \left( \frac k{k_0}\right) ^{1/\left( C+1\right)
}-4\right] }\right\}  \nonumber \\
&&\times \left\{ \left( \frac k{k_0}\right) ^{1/\left( C+1\right) }
+2C-\sqrt{\left(\frac k{k_0}\right)^{1/\left(C+1\right)}
\left[\left(\frac k{k_0}\right)^{1/\left(C+1\right)}-4\right] }\right\}
\, .
\label{eq:Dk0}
\end{eqnarray}
Plot for $\Delta(k)$ and $\delta=0$ is presented in the left part of
Fig.~\ref{fig:d0Dk}.
The range of scales plotted was chosen to allow observation of features
at lowest scales. The logarithmic plot of a proper range of scales is presented
in the right part of this figure, where in the vertical axis is plotted the
logarithm of squared scalar amplitudes given by,
\begin{eqnarray}
\ln\frac{A_S(k)^2}{A_0^2} &=&-\ln\left(\frac12\left\{ \left( \frac
k{k_0} \right) ^{1/\left( C+1\right) }-2
-\sqrt{\left( \frac k{k_0}\right) ^{1/\left(C+1\right) }
\left[\left(\frac k{k_0}\right)^{1/\left(C+1\right)}-4\right] }
\right\}\right)  \nonumber \\
&&+\ \frac1{8\left(C+1\right)}\left\{ \left(\frac
k{k_0}\right)^{1/\left(C+1\right)}-2
-\sqrt{\left( \frac k{k_0}\right) ^{1/\left( C+1\right)
}\left[ \left( \frac k{k_0}\right) ^{1/\left( C+1\right)
}-4\right] }\right\} \nonumber \\
&&\times \left\{\left(\frac
k{k_0}\right)^{1/\left(C+1\right)} -2+4C
-\sqrt{\left(\frac k{k_0}\right)^{1/\left(C+1\right)}
\left[\left(\frac k {k_0}\right)^{1/\left(C+1\right)}-4\right] }
\right\} . \label{eq:Ak}
\end{eqnarray}
The value for $A_0$ must be chosen taking into account the
observational constrain $A_S^2\sim 10^{-5}$ given by COBE
measurements.

From these figures it is observable that the scalar index could be
regarded as scale independent for relevant to measurements scales.
The corresponding constant value is $n_S=(C+2)/(C+1)\simeq 4.7$,
which is too far from values allowed by theory and experiments. It
seems to be not possible an inflationary model to exist such that
the tensorial spectrum generated in its framework will be of
Harrison-Zeldovich type and the corresponding scalar index will be
scale-dependent. Therefore, to this order and considering
Eqs.~(\ref{eq:rnT}) and (\ref{eq:StLythNt}) with $\delta=0$, it
can be concluded that any model with a scale-dependent scalar
index matching observations will give a nonzero tensorial
contribution to the CMB spectrum.

Note that, while applying criterion (\ref{eq:mepsConds}) to the mirror
image with respect to the ordinate axis of solution (\ref{eq:Vdelta0}), the
same result is obtained.

\subsubsection{SLIP solution for negative tensorial index, $\delta<0$}
\label{subsec:general}

We shall find the inflationary potentials producing perturbations that, to
next-to-leading order, are characterized by constant and negative tensorial
index and scale-dependent scalar index.
Since for $\delta<0$, SLIP equations are not explicitly solvable in terms
of $\phi$, we must look for a parametric expression for the inflaton
potential.

\begin{figure}[h]
\centerline{\psfig{file=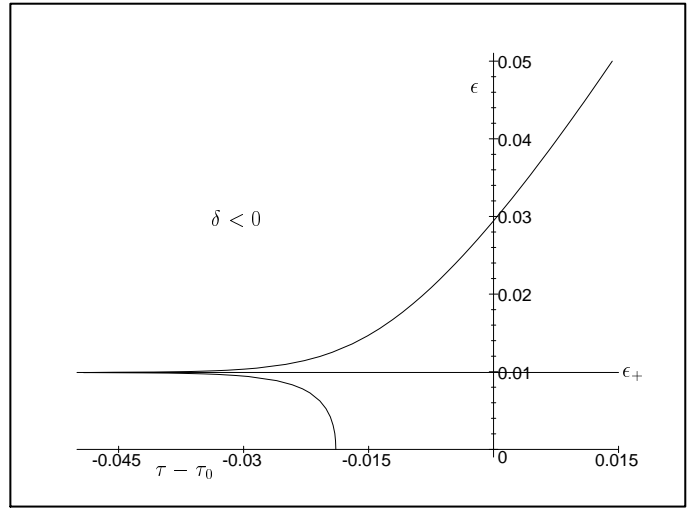,width=9.5cm}}
\caption{Solution of the first order equation for $\delta<0$.
Each curve branch that goes from left to right is a solution with
different initial values.}
\label{fig:dlt0et}
\end{figure}

\begin{figure}[h]
\centerline{\psfig{file=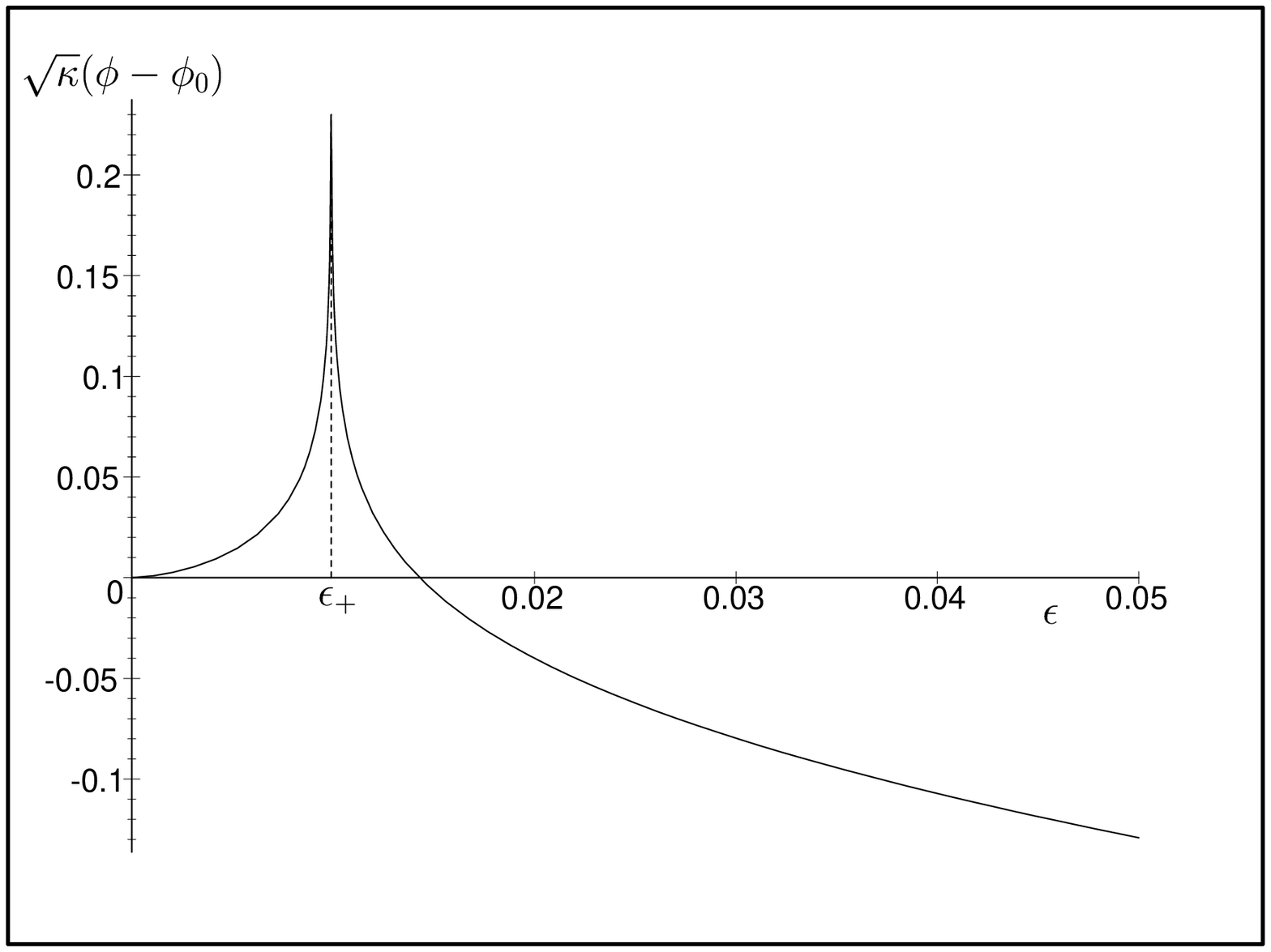,width=9.5cm}
\psfig{file=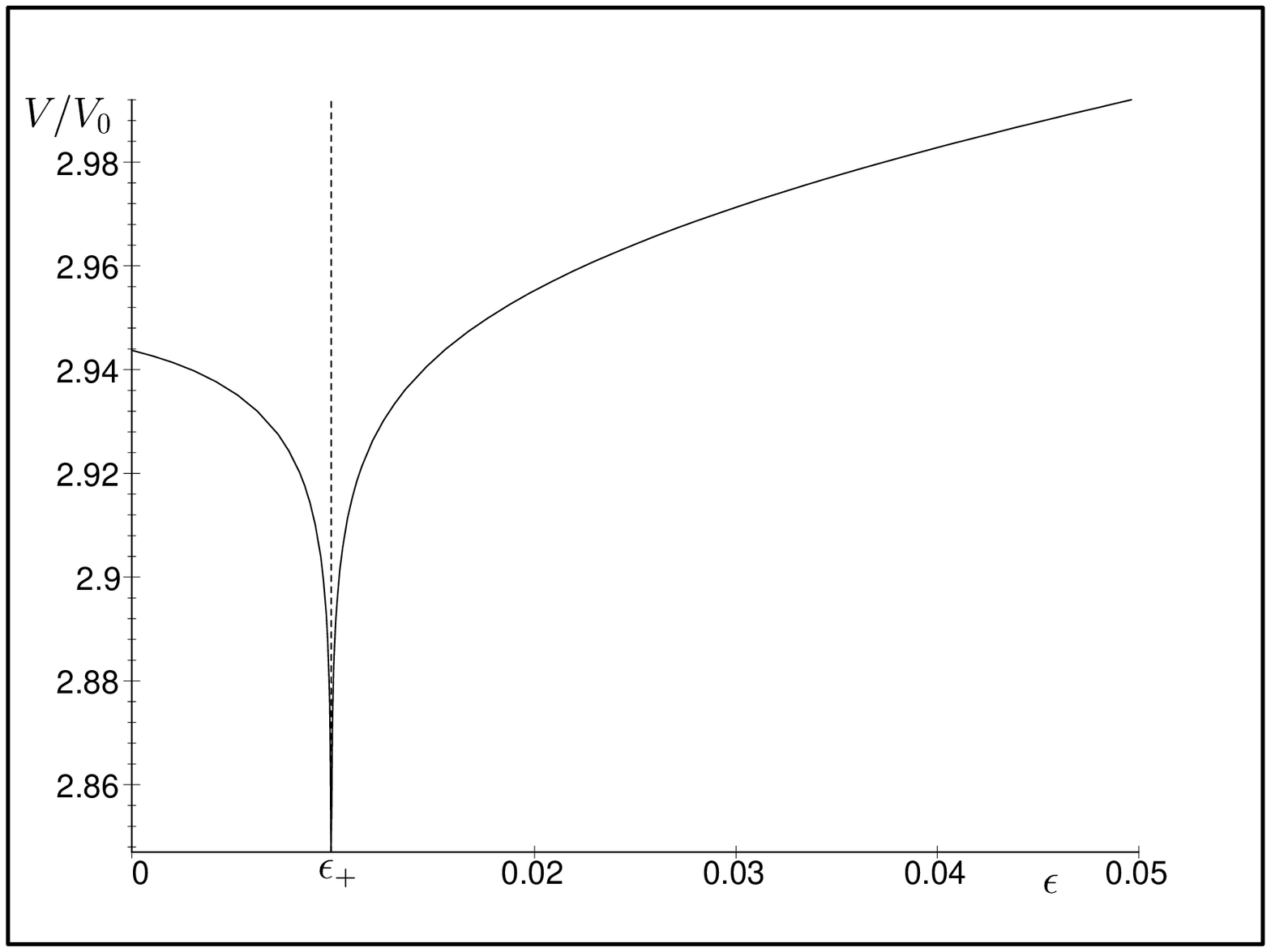,width=9.5cm}}
\caption{Inflaton and its potential as functions of the first
slow-roll parameter for $\delta=-0.01$.}
\label{fig:dlt0pVe}
\end{figure}

The solution of Eq.~(\ref{eq:MSch2}) for $\delta<0$ is \cite{Nos}
\begin{equation}
\exp(\tau-\tau_0)=\left|\epsilon ^{2}+\epsilon
+\delta\right|^{C+1}
\left|\frac{2\epsilon+1+\sqrt{1-4\delta}}
{2\epsilon+1-\sqrt{1-4\delta}}\right|^{\frac{C+1}{\sqrt{1-4\delta}}}
\,.
\label{eq:FO3}
\end{equation}
The three branches corresponding to this expression are plotted in
Fig.~\ref{fig:dlt0et}.
It is observed that, along with stationary solution
\begin{equation}
\epsilon_+ = -\frac{1}{2} + \frac{1}{2}\sqrt{1-4\delta}\, ,
\label{eq:StatSol}
\end{equation}
solutions can increase unbounded or decrease bounded by value
$\epsilon=0$.
Regarding that $\tau$ and cosmic time have opposite signs, there is some time
interval where $\epsilon\simeq\epsilon_+$, i.e., the solution asymptotically
behaves like that of power-law. After integrating
Eq.~(\ref{eq:Phi}) for $\delta<0$ and inserting solution
(\ref{eq:FO3}) in expression (\ref{eq:PotentialE}), the parametric
potential is given by
\begin{equation}
V(\phi)= \left\{
\begin{array}{rcl}
\phi(\epsilon)&=&\frac{2(C+1)}{\sqrt{\kappa}}\frac{1}{\sqrt{1-4\delta}}
\left[-\sqrt{1+\sqrt{1-4\delta}}\arctan\left(\frac{\sqrt{2\epsilon}}
{\sqrt{1+\sqrt{1-4\delta}}}\right)\right.\\
&& + \left. \frac{1}{2} \sqrt{-1+\sqrt{1-4\delta}}
\ln\left|\frac{\sqrt{2\epsilon}+\sqrt{-1+\sqrt{1-4\delta}}}
{\sqrt{2\epsilon}-\sqrt{-1+\sqrt{1-4\delta}}}\right|\right]
+ \phi_0 \, ,  \\
&& \\
V(\epsilon)&=&V_0(3-\epsilon)
\left|\epsilon ^{2}+\epsilon +\delta\right|^{C+1}
\left|\frac{2\epsilon+1+\sqrt{1-4\delta}}
{2\epsilon+1-\sqrt{1-4\delta}}\right|^\frac{C+1}{\sqrt{1-4\delta}}
\, .
\end{array}
\right.
\label{eq:Sol4}
\end{equation}
The inflaton and corresponding potential as functions of $\epsilon$ are
respectively plotted in Fig.~\ref{fig:dlt0pVe}.
Note that the graph of the potential has also a
maximum around
\[
\epsilon^*\equiv\frac{\left[(6C+5)+\sqrt{(6C+5)^2-4\delta(2C+3)}\right]}
{(4C+6)}\simeq0.4
\]
for $\delta=-0.01$, not shown in the figure in order to observe
the details for small values of $\epsilon$. Hence, the analysis
for $\delta<0$ should be done for three intervals of $\epsilon$,
namely, $I_1=[0,\epsilon_+)$, $I_2=(\epsilon_+, \epsilon^*]$, and
$I_3=(\epsilon^*, 1)$.

\begin{figure}[h]
\centerline{\psfig{file=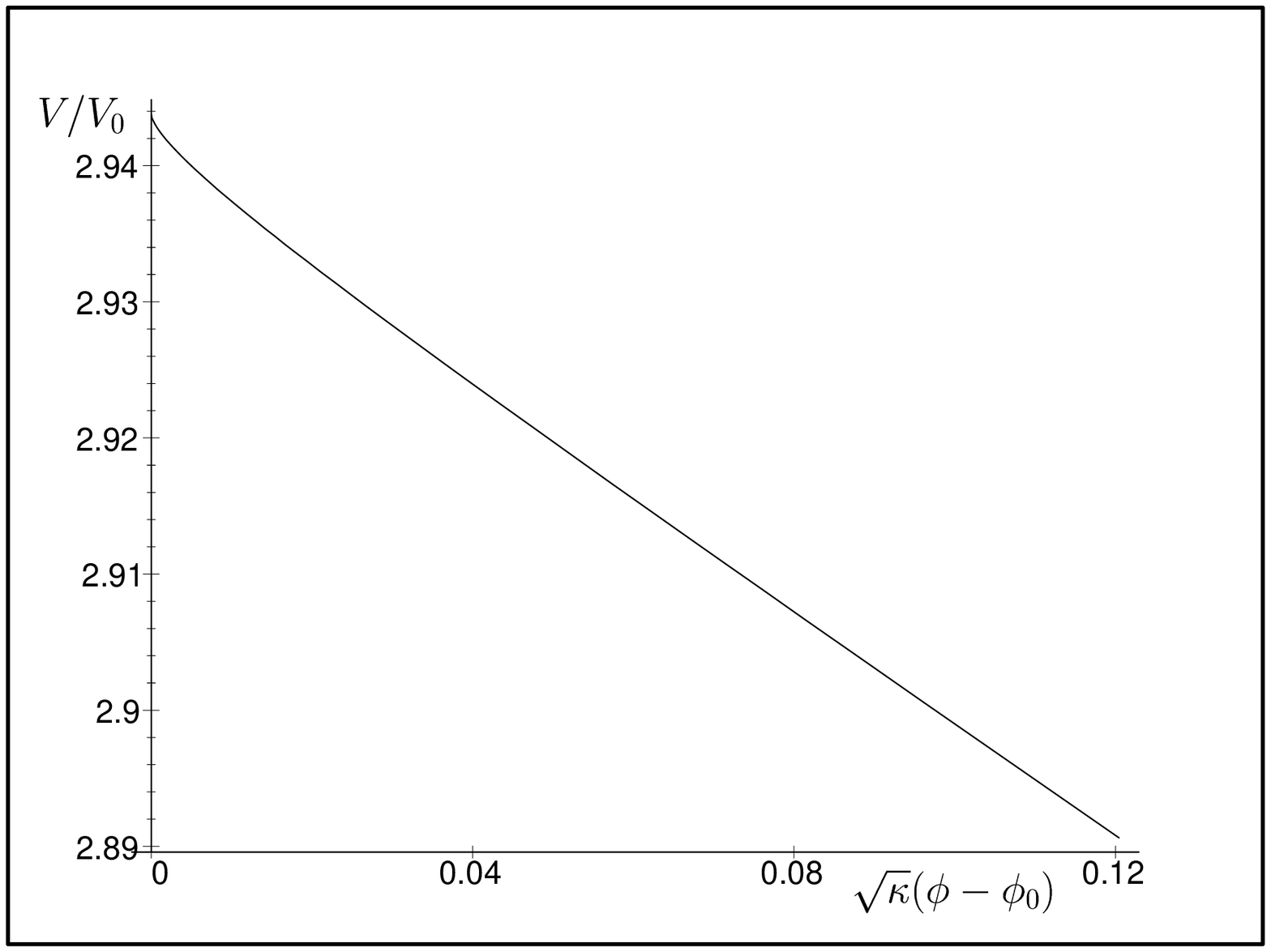,width=9.5cm}
\psfig{file=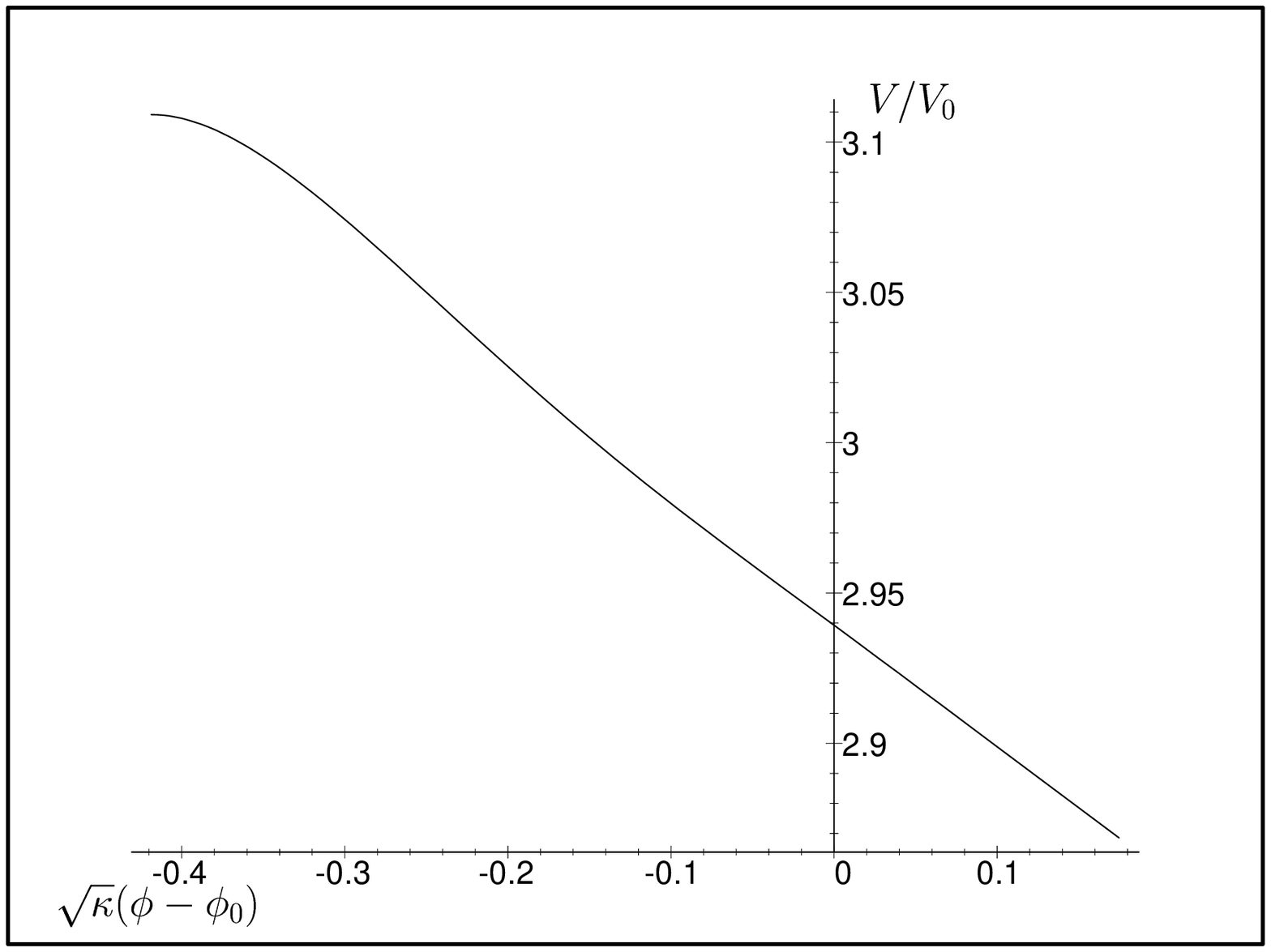,width=9.5cm}}
\caption{Consistent SLIP solutions for $\epsilon\in I_1$ and
$\epsilon\in I_2$, respectively, with $\delta=-0.01$.}
\label{fig:dlt0Vp}
\end{figure}

Making use of criterion (\ref{eq:epsConds}), consistent SLIP
solutions are determined to exist in the intervals $I_1$ and
$I_2$. Corresponding plots are presented in
Fig.~\ref{fig:dlt0Vp}.
Here, the value of the constant $V_0$ could be chosen to make the
potential flat enough for conditions of successful inflation
to be satisfied. For $\epsilon \in I_3$, the solution fails to
fulfill criterion (\ref{eq:epsConds}).

The main difference between both solutions is the curvature of the
inflaton potential near the origin, but is not a trivial one. For the same
parametric expression of the potential we have two rather
different realization of inflation depending not only on initial
conditions for $\phi$ but also for $\dot{\phi}$ (i.e., depending
on the initial conditions for $\epsilon$). In the case corresponding
to the left graph on Fig.~\ref{fig:dlt0Vp},  the
term in Eq.~(\ref{eq:mass}) given by the first derivative of the potential
will generically dominate during the inflationary epoch
meanwhile, in the remaining case, the evolution of $\phi$ will be
dominated by the friction term due to Universe expansion,
$3H\dot{\phi}$.

\begin{figure}[h]
\centerline{\psfig{file=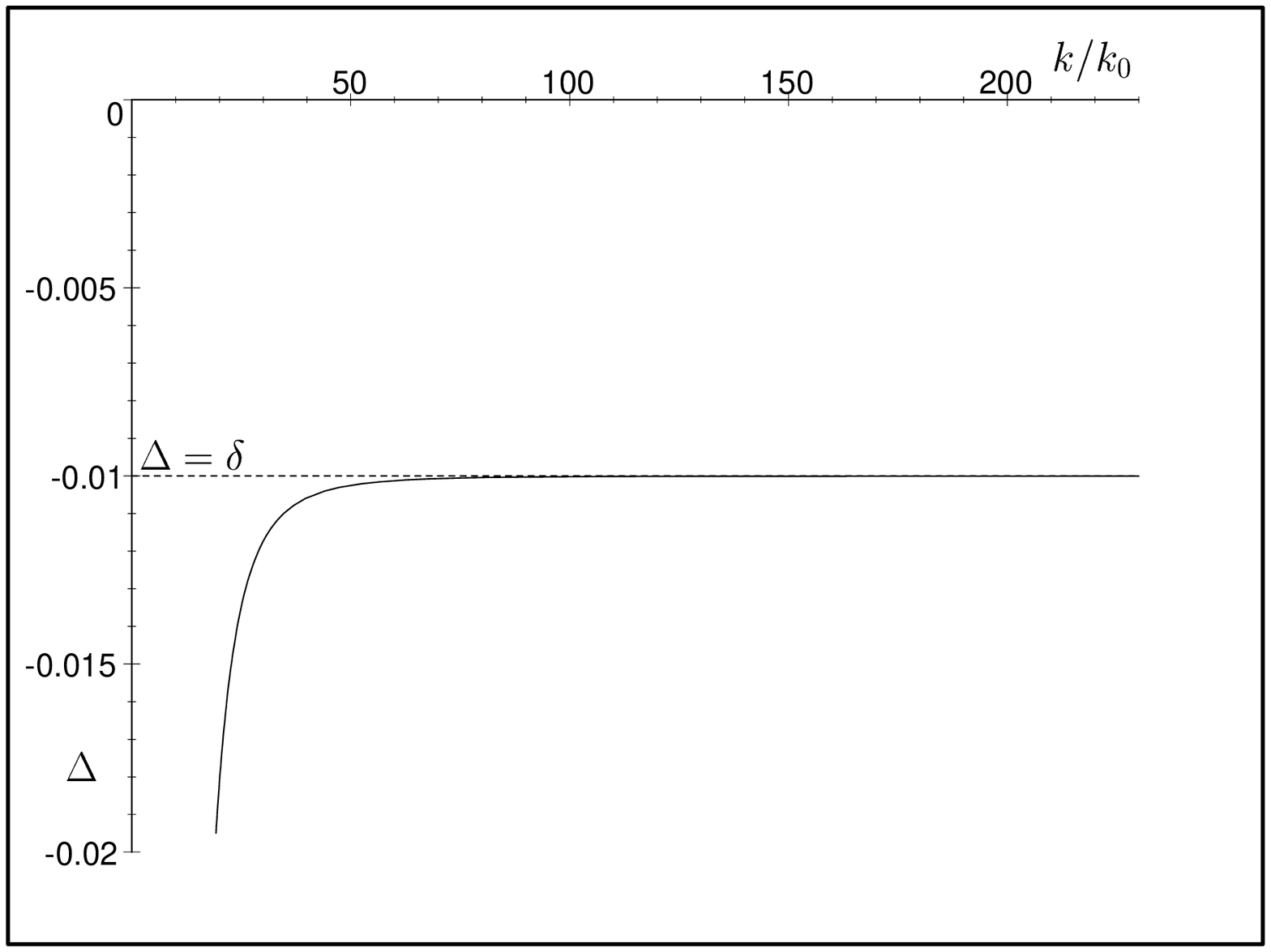,width=9.5cm}
            \psfig{file=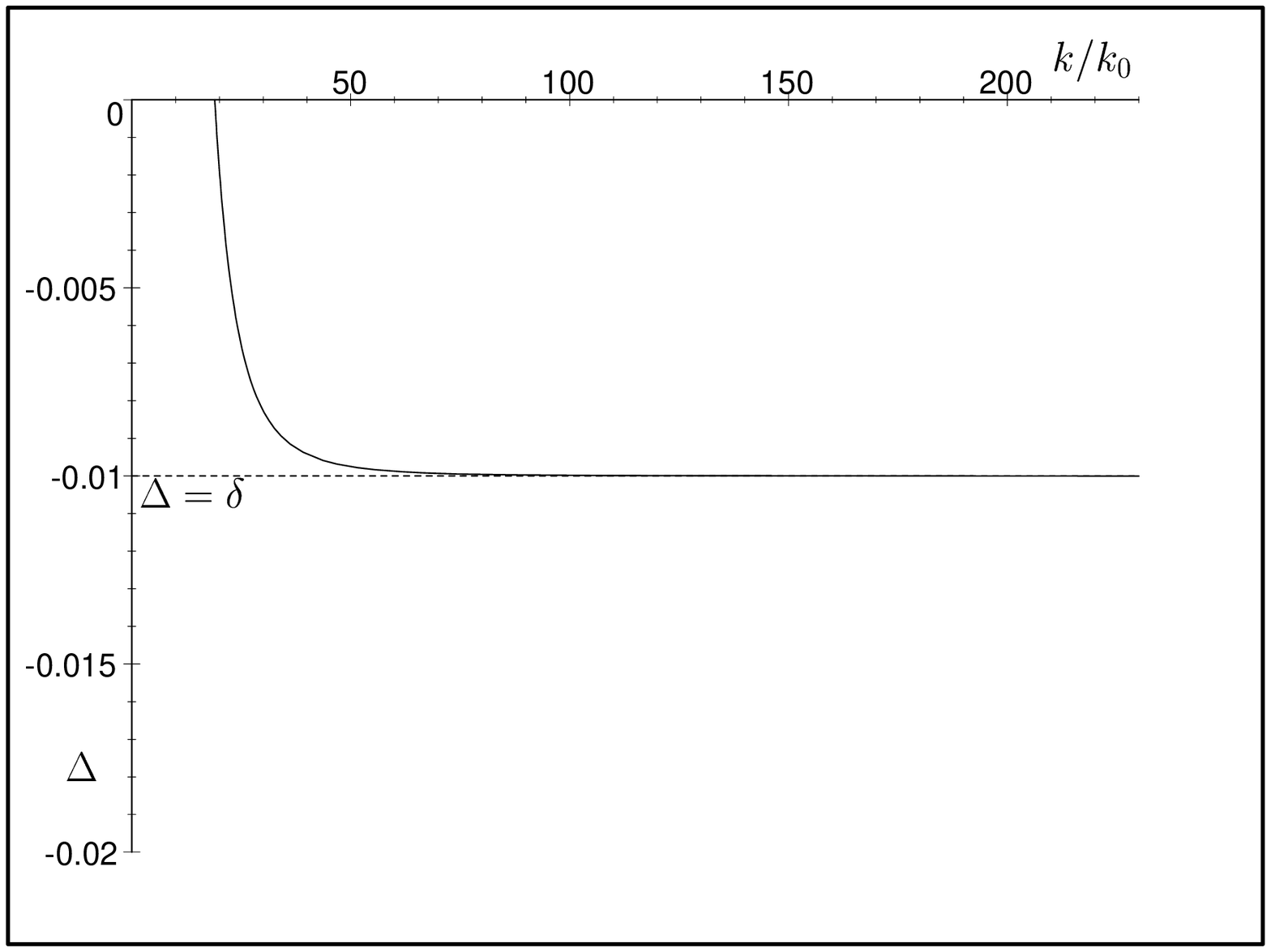,width=9.5cm}}
\caption{$\Delta$ as a function of the comoving number for
$\epsilon\in I_1$ and $\epsilon\in I_2$, respectively, and $\delta=-0.01$.}
\label{fig:dlt0Dk}
\end{figure}

For $\delta<0$, after integration of Eq.~(\ref{eq:ke}) we have that
\begin{equation}
k= k_0\left|\epsilon ^{2}+\epsilon
+\delta\right|^\frac{C+1}{2}
\left|\frac{2\epsilon+1+\sqrt{1-4\delta}}
{2\epsilon+1-\sqrt{1-4\delta}}\right|^{\frac{3(C+1)}{2\sqrt{1-4\delta}}}
\,,
\label{eq:k3}
\end{equation}
where $k_0$ is the integration constant. With $\Delta(\epsilon)$ given by
Eq.~(\ref{eq:De}), parametric plots for $\Delta(k)$ corresponding to
$\epsilon\in I_1$ and $\epsilon\in I_2$ are presented in
Fig.~\ref{fig:dlt0Dk}.
Again, the scale range was chosen to allow
details observation at lowest scales. Observe that in
both cases the present error for $n_S$ (e.g., $n_S=0.99\pm0.09$ in
Refs.~\cite{observ2}) can mask the scale dependence at large
$k^{-1}$.

After integrating Eq.~(\ref{eq:Ase}) for $\delta<0$, the logarithm of squared
scalar amplitudes is given by
\begin{eqnarray}
\ln \frac{A_S(\epsilon )^2}{A_0^2} &=&\frac{\delta \,C-1-C}{C+1}\ln \epsilon
+\delta \left( C+1\right) \ln \left| \epsilon ^2+\epsilon +\delta \right|
\nonumber \\
&&+\ \frac{3\delta \left( C+1\right) }{\sqrt{1-4\,\delta }}\ln \left| \frac{%
2\,\epsilon +1+\sqrt{1-4\,\delta }}{2\,\epsilon +1-\sqrt{1-4\,\delta }}%
\right| \nonumber \\
&&+\ \frac{\epsilon ^3+2C\epsilon ^2+2C\delta }{2\left( C+1\right)
\epsilon \,}  \, .
\label{eq:dlt0Ae}
\end{eqnarray}
\begin{figure}[h]
\centerline{\psfig{file=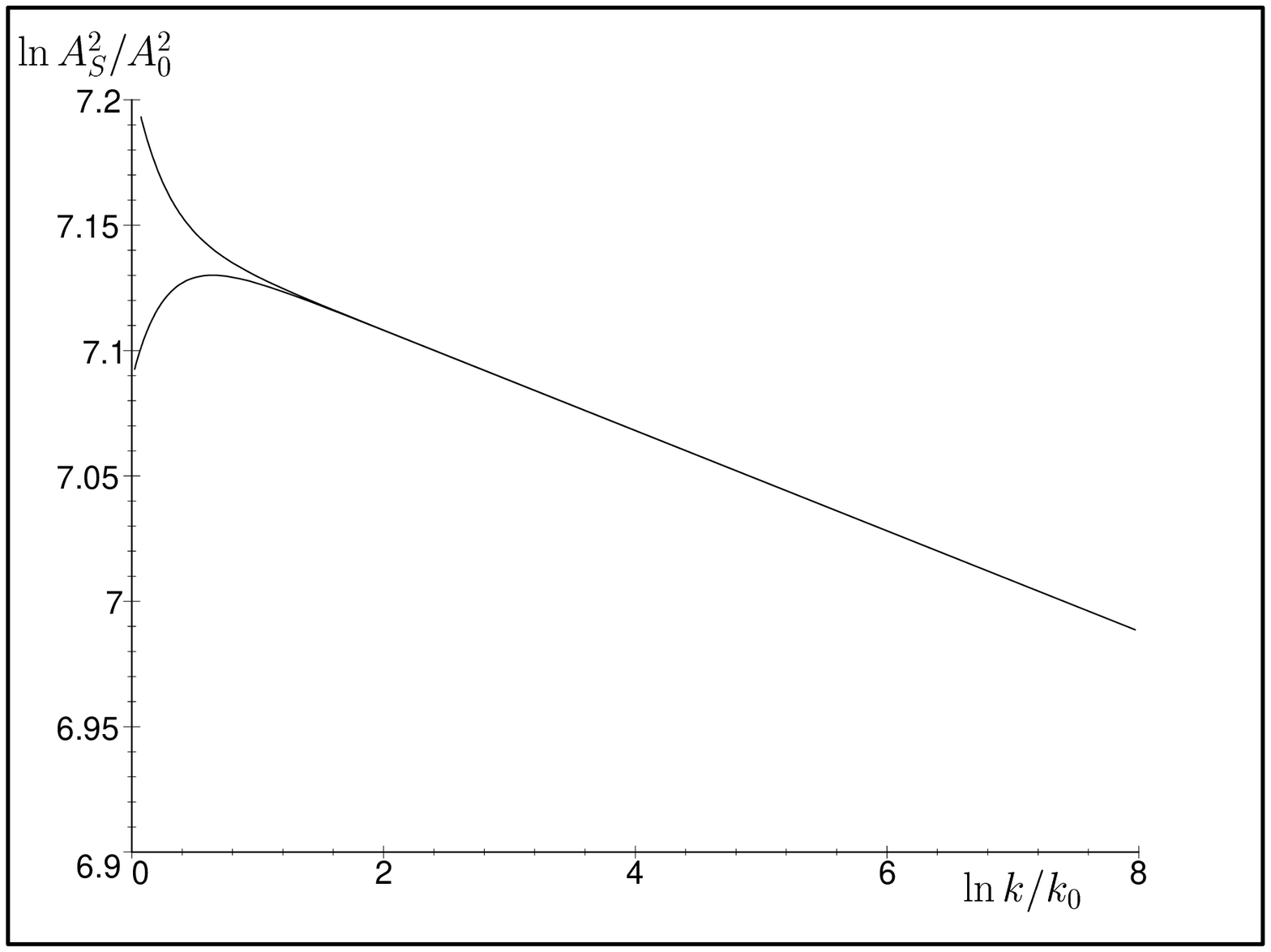,width=9.5cm}}
\caption{$\ln A^2_S$ as a function of the $\ln k$ for $\delta=-0.01$
and for $\epsilon\in I_1$ (lower branch) and $\epsilon\in I_2$ (upper branch).}
\label{fig:dlt0lnAlnk}
\end{figure}
As observed in Fig~\ref{fig:dlt0lnAlnk}, where the parametric
plots of scalar amplitudes for $\epsilon\in I_1$ and $\epsilon\in
I_2$ is presented in the same figure, differences are almost
impossible to note when the full range of scales is considered.
Differences arise at large angular scales which could be out of
reach for measurements. It means that, from the observational point
of view, there could be not differences between these two
realizations of inflation if the scales where differences arise
are not probed or the resolution is not high enough to detect the
scale dependence. These scales are precisely those where higher
energies physics could leave an imprint.

Recalling the behavior of solution (\ref{eq:FO4}) for small
$\tau$, i.e., large $t$ (see Fig.~\ref{fig:dlt0et}) and taking the
limit $\epsilon\rightarrow\epsilon_+$ of Eqs.~(\ref{eq:De}) and
(\ref{eq:k3}) it is found out that for small $k^{-1}$,
$\Delta\simeq\delta$ (see also Fig.~\ref{fig:dlt0Dk}). That the
scale interval where this behavior is observed corresponds to
sufficient large number of e-folds to solve the Standard Model problems
is provided by the asymptotic behavior of $\epsilon$ near the
value $\epsilon_+$ .

These models have the desired feature of an almost negligible
$\delta=\rm const.$ and the scalar index being nearly constant in
a wide range of scales, with the possibility of choosing values
that can accurately match current observations. In fact, the value
$\delta=-0.01$ chosen in the figures of this section corresponds
(in the limit $\epsilon\rightarrow\epsilon_+$)
to $n_S=0.98$ compatible with values given in
Refs.~\cite{observ2}. Any other constant value for $n_S$ arising from
analysis similar to that of
Refs.~\cite{observ2,Zalda,Kinney,Covi,Tegmark}, except for blue spectra, 
can also be fitted with
models given by Eq.~(\ref{eq:Sol4}). The weak scale-dependence of $n_S$ 
obtained is
in good agreement with results in Ref.~\cite{Steen}. Furthermore, 
the above mentioned value of
$\delta$ approximately corresponds to $r\simeq 0.12$ which is
greater than $r^*=0.1$, the value given in Ref.~\cite{LLBook} as
the lower limit for $r$ to be detectable with a 95-percent
confidence regarding the error reported in Ref.~\cite{Zalda} as the estimate
for Planck measurements. Lower values for $r$ can be appropriately taken
into account.

Because models given by Eq.~(\ref{eq:Sol4}) do not
have a graceful exit to the Standard Model stage of
Universe evolution ($\epsilon$ converges to $\epsilon_+$ not to
$1$), $\phi$ must be regarded here as the dominant scalar field in
a hybrid scenario with $\epsilon_+$ being the value corresponding
to the critical value of $\phi$ near which the false
vacuum becomes unstable and the multiple scalar fields roll to the
true potential minimum \cite{LLBook}.

Finally, all of the statements done in this section with regards to solution
(\ref{eq:Sol4}) are also valid after changing $\phi(\epsilon)$ by 
$-\phi(\epsilon)$ and using criterion (\ref{eq:mepsConds}).

\section{Conclusions}
\label{sec:conclu}

We presented a version of Stewart-Lyth inverse problem
using the first slow-roll parameter as the basic variable in the
procedure of finding the inflaton potential. That allows us to analyze
the solutions in the range of this parameter where inflation is feasible.
A criterion was introduced to check for solutions consistent with the 
assumptions underlying the
derivation of the Stewart-Lyth inverse problem equations.

It was shown that expressions related to Stewart-Lyth inverse
problem can be used to determine inflationary models corresponding
to given observations. We proved that
power-law inflation is a trivial solution of this problem when
constant spectral indices are used as input in related equations.
Next-to-leading-order in the slow-roll expansion makes possible
to consider more general scenarios where slow-roll parameters can
slowly vary with time. In a near future, these scenarios could be more
realistic than common assumption regarding slow-roll parameters as
constants during inflation.

Looking for a potential generating the primordial perturbations
able to grow into CMB anisotropies and matching current and
future observations, we solved the Stewart-Lyth inverse problem
with constant tensorial index as input. Inflationary
models were found which, unlike power-law inflation, yield scalar modes
characterized by a scale-dependent index. For negative tensorial index,
solutions were given as an expression depending on the first
slow-roll parameter.

The special case of a Harrison-Zeldovich spectrum of tensorial
perturbations, i.e., constant amplitudes of gravitational waves, 
is ruled out by comparison of our results with current
observations. It means that it seems to not exist inflationary
models with scale dependent scalar index and null tensorial index.
Hence, for any model exhibiting some degree of scale dependence
of the index of curvature perturbations it must be expected a nonzero
contribution of primordial gravitational waves to the amplitudes of
the CMB spectrum.

Potentials obtained for strictly negative tensorial index fulfill
the conditions for successful
inflation. Evolution of the scalar field given by these potentials
can be considered as the dynamical element in a hybrid
inflation scenario, the value of the inflaton corresponding to
power-law solution acting like the instability value for the
false vacuum.

These models can be used as assumption on the origin of
primordial perturbations to test for scale dependence of the 
scalar index.
Using them to fit inflationary perturbations is almost as easy as using
power-law inflation. Only one
parameter, the constant tensorial index, need to be fitted. If CMB
polarization fails to give a value of the tensor-scalar ratio
greater than the threshold value $r^*=0.1$, then the tensorial
index must be fitted in the interval $(-0.02,0)$. Otherwise, if
some value for the tensor-scalar ratio is measured, then an
approximated value for the tensorial index can be estimated to
serve like pivot value for the fitting procedure.

We would like to stress that if any of the potentials
here presented makes possible to reach an overall good fit for CMB
anisotropies and to detect scale dependence of the scalar index
from the next generation of observations, the conclusion to be
drawn is that the actual inflaton potential is similar in the
probed scale to the used one. In turn, if the quality of the
overall fit is not improved compared to the result obtained using
power-law inflation, either the scale dependence of the scalar
index is practically negligible, or the information on the
scale dependence of the tensorial index is fundamental in order to account
for the features in the CMB anisotropies.

The spectra of scalar perturbations produced by these models differ
from those of power-law inflation at scales corresponding to earlier times
in the Universe evolution. Thus, if while increasing the quality
of observations a good fit using any of these potentials is
achieved, it could give some hints about physics taking place at
very high energies.

\acknowledgments

This research is supported in part by the CONACyT grant 32138-E.
The work of CATE was also partially founded by the Sistema Nacional de
Investigadores (SNI) and CINVESTAV. We want to thank Andrew Liddle and 
Dominik Schwarz for helpful discussions.

\end{document}